\documentclass[%
 pre,
rsi,%
 amsmath,amssymb,
 reprint,%
]{revtex4-2}

\usepackage{xcolor}
\usepackage{xfrac}
\usepackage{empheq}
\usepackage{stmaryrd}
\usepackage{leftidx}
\usepackage{ esint }
\usepackage{amsmath}
\usepackage{amssymb}
\usepackage{mathtools}
\usepackage{graphicx}
\usepackage{dcolumn}
\usepackage{bm}

\begin{document}

\title[Sensitivity of long periodic orbits]{Sensitivity of long periodic orbits of chaotic systems}

\author{D. Lasagna}
 \altaffiliation[]{Faculty of Engineering and Physical Sciences, University of Southampton, United Kingdom}


\date{\today}

\begin{abstract}
The properties of long, numerically-determined periodic orbits of two low-dimensional chaotic systems, the Lorenz equations and the Kuramoto-Sivashinsky system in a minimal-domain configuration, are examined. The primary question is to establish whether the sensitivity of period averaged quantities with respect to parameter perturbations computed over long orbits can be used as a sufficiently good proxy for the response of the chaotic state to finite-amplitude parameter perturbations. To address this question, an inventory of thousands of orbits at least two orders of magnitude longer than the shortest admissible cycles is constructed. The expectation of period averages, Floquet exponents and sensitivities over such set is then obtained. It is shown that all these quantities converge to a limiting value as the orbit period is increased. However, while period averages and Floquet exponents appear to converge to analogous quantities computed from chaotic trajectories, the limiting value of the sensitivity is not necessarily consistent with the response of the chaotic state, similar to observations made with other shadowing algorithms.

\end{abstract}

\pacs{Valid PACS appear here}
\maketitle

\section{Introduction}\label{sec:introduction}
Evidence has been offered in recent years \cite{Viswanath:2007cy,Chandler:2013fi,Willis:2016fq,Budanur:2017jc} that temporally-recurrent invariant solutions of the Navier-Stokes equations  -- unstable periodic orbits --  may provide a constitutive skeleton organizing spatiotemporal dynamics of turbulent shear flows in canonical geometries. 
Motivated by these advances, we have recently suggested \cite{Lasagna:2017tz} how unstable periodic orbits may be used to design control strategies for shear flows, rather than serving as a tool to rationalise turbulence dynamics. In particular, \textcolor{black}{in Ref.~\cite{Lasagna:2017tz},} we have specialized adjoint methods for time-periodic systems \cite{Hwang:2008bv,Giannetti:2010jb,Meliga:2016fjb,Giannetti:2019do} to unstable periodic orbits. We have shown that enforcing periodicity conditions on the adjoint problem, justified by the peculiar topology of these trajectories, prevents the growth of exponential instabilities that would otherwise feature prominently in the solution of the adjoint equations \cite{2000TellA..52..523L,Luchini:2002dj}. 

Operationally, the approach provides the sensitivity of period averaged quantities with respect to small perturbations of variables parametrising the equations of motion. Geometrically, small parameter perturbations can be pictured as producing smooth, global state-space deformations of the unstable periodic orbits supporting and shaping the attractor, as opposed to causing exponential divergence. The approach is a special case of shadowing theory ideas \cite{Bowen:1975hc, Ruelle:1999bm, Administrator:2018wr}, recently introduced in the context of sensitivity analysis of chaotic systems \cite{Wang:2013cx,Wang:2014hu,Ni:2017bs,Shawki:2018vz}.  


In principle, complete knowledge of the short, fundamental cycles should suffice to compute ergodic averages using cycle averaging formulae \cite{Eckhardt:1994dv,Cvitanovic:1995gu,Artuso:1999ez}. Of relevance for our original motivation is that a formalism that relies on the sensitivity of such cycles to compute the sensitivity of ergodic averages was proposed in Refs. \cite{Kazantsev:1998wb,Kazantsev:2001fx}. Obtaining all short cycles up to a given topological length, \textcolor{black}{i.e.~low-period orbits identified by a short symbol sequence \citep{Cvitanovic:1995gu,Dong:2014eva}}, may be practical for low-dimensional systems (see e.g. Ref.~\cite{Viswanath:2003gb}). However, this step has proven more challenging for turbulent shear flows \cite{Chandler:2013fi,Lucas:2015gt}, given well documented difficulties in locating invariant solutions \cite{vanVeen:2019bi,Lucas:2020to}. This issue is particularly relevant, since the quality of cycle averaging predictions using incomplete hierarchies is as good as the most important orbit that one fails to locate \cite{Budanur:2015ic}. 


In light of such issues, we explore in this paper a heuristic approach whereby available computational resources are spent to locate \textcolor{black}{one or a few} periodic orbits, with sufficiently long period for them to span a good fraction of the attractor. The open question is whether sensitivities of period averaged quantities computed over a long orbit can be useful approximations for the response of ergodic averages to parameter perturbations. Like period averages, sensitivities computed over different periodic orbits vary from cycle to cycle. However, it is known that \textcolor{black}{period averaged quantities calculated from} long periodic orbits converges to a defined value when the period increase \cite{Zaks:2010ex}. Evidence showing a similar convergence for sensitivities of period averages is currently not available in the literature and would provide initial support for the above heuristic. This approach is, admittedly, guided more by empiricism rather than by a solid theoretical basis. Hence, the aim of this paper is to make a first step in exploring its viability. We resort, by necessity, to low-dimensional systems,  where obtaining a sizeable inventory of long periodic orbits is feasible. We consider the Lorenz equations at standard parameters \cite{Lorenz:1963tf} and a small-domain Kuramoto-Sivashinsky system in the antisymmetric subspace \cite{Christiansen:1997gd,Lan:2008kg}.

One remark is in order. It is true that some long periodic orbits might not provide good approximations, for instance orbits close to bifurcation or orbits visiting certain areas of the attractor where the response is particularly large. As an illustrative example, in Ref.~\cite{Lasagna:2017tz}, long periodic orbits of the Lorenz equations passing close to the origin were found that have extremal value of the sensitivity. Hence, to develop a quantitative understanding of the potential impact of such extremal orbits, we locate thousands of long periodic orbits by converging near recurrences explored by the chaotic flow and report sensitivity predictions of these periodic orbits in a statistical manner. The analysis of this ensemble is clearly bound to suffer of some form of survivor bias. For instance, Newton-Raphson techniques often fail to converge when the orbit shadowed by the near recurrence event possesses multipliers close to the unit circle \cite{Sanchez:2004cs}. Nevertheless, this bias does not prevent us to complete the original task, which is to analyse the properties of \textit{typical} periodic orbits that may be found numerically rather than the properties of all admissible cycles.

The rest of this paper is organized as follows. In section \ref{eq:preliminaries} we lay down the general notation. This is followed by section \ref{sec:floquet-analysis}, where we recall fundamental elements of Floquet theory for the linear stability of periodic orbits \cite{Farkas:2013gm,Moore:2006kj}. In section \ref{sec:sensitivity-analysis} we recall the sensitivity technique for unstable periodic orbits originally introduced in Ref.~[\onlinecite{Lasagna:2017tz}]. With this introductory material, in section \ref{sec:forward-sensitivity-floquet} we establish the connection between Floquet stability and sensitivity of periodic orbits. This analysis illustrates more clearly the inevitable effects of bifurcations that periodic orbits undergo as parameters are varied and supports the interpretation of numerical results, reported in section \ref{sec:numerical-results}.






\section{Preliminaries}\label{eq:preliminaries}
We consider dissipative dynamical systems of the form
\begin{equation}\label{eq:system}
	\frac{\mathrm{d}\mathbf{u}(t)}{\mathrm{d}t} = \mathbf{f}(\mathbf{u}(t), \alpha),
\end{equation}
governing the evolution of the state vector $\mathbf{u}(t) \in \mathbb{R}^{n}$, with $t$ being time. 
We restrict the attention to problems where the nonlinear vector field $\mathbf{f}$ depends only on one parameter $\alpha \in \mathbb{R}$, in regimes for which chaotic solutions are observed for typical initial conditions. For systems that depend on multiple parameters, the sensitivity of time averaged quantities can be analysed over each parameter independently, or with an adjoint approach. We denote with $\mathbf{f}_\mathbf{u}(t) \in \mathbb{R}^{n\times n}$ the stability matrix
\begin{equation}
	\mathbf{f}_\mathbf{u}(t) = \mathbf{f}_\mathbf{u}(\mathbf{u}(t), \alpha) = \frac{\partial \mathbf{f}(\mathbf{u}(t), \alpha)}{\partial \mathbf{u}}.
\end{equation}
and define the Jacobian matrices $\mathbf{M}(t, \tau)\in \mathbb{R}^{n\times n}$, $t \ge \tau$, satisfying the initial value problem
\begin{equation}\label{eq:jacobian-matrices}
	\frac{\mathrm{d}\mathbf{M}(t, \tau)}{\mathrm{d}t} = \mathbf{f}_\mathbf{u}(t) \mathbf{M}(t, \tau), \quad \mathbf{M}(\tau, \tau) = \mathbf{I},
\end{equation}
with $\mathbf{I}$ the identity matrix.
The dot product of two vectors is denoted with $\mathbf{a}^\top\cdot\mathbf{b}$, with the script ${}^\top$ indicating transposition of vectors and matrices.

We focus on periodic solutions of (\ref{eq:system}), satisfying \mbox{$\mathbf{u}(t + T) = \mathbf{u}(t)$} for an unknown period $T$ that is not set a priori, but depends implicitly on the parameter $\alpha$. We shall thus consider the space of smooth periodic functions
\begin{equation}
	\mathcal{P}_T = \{f(t) : \mathbb{R} \mapsto \mathbb{R}, f(t)=f(t+T)\},
\end{equation}
parametrized by $T$, and extend this space to vector-valued functions, denoted with $\mathcal{P}_T^n$. We will make use of the norm $\|\cdot\|_{\mathcal{P}_T^n}$ induced by the inner product 
\begin{equation}\label{eq:inner-product}
	\left\llbracket\mathbf{v}(t), \mathbf{w}(t) \right\rrbracket = \frac{1}{T} \int_0^{T} \mathbf{v}(t)^\top\cdot\mathbf{w}(t)\,\mathrm{d}t,
\end{equation}
for any two vector-valued functions $\mathbf{v}(t)$ and $\mathbf{w}(t)$ in $\mathcal{P}_T^n$.

\section{Elements of Floquet theory}\label{sec:floquet-analysis}
In this section, we briefly recall some elements of Floquet theory \cite{Guckenheimer:up,Moore:2006kj} and define the direct and adjoint Floquet eigenfunctions. These functions are used in section \ref{sec:floquet-analysis} as invariant subspaces in which the solution of the sensitivity problem can be conveniently expanded, shedding light on the relation between stability and sensitivity of period averages.

For an arbitrary point on a periodic orbit, the eigenvalue decomposition of the monodromy matrix
\begin{equation}\label{eq:eigenvalues-of-monodromy}
	\mathbf{M}(T, 0)\mathbf{e}_k(0) = \mu_k\mathbf{e}_k(0),\quad k = 1, \ldots, n,
\end{equation}  
produces the Floquet multipliers $\mu_k$ and the associated right eigenvectors $\mathbf{e}_k(0)$. To simplify the notation and the analysis of section \ref{sec:forward-sensitivity-floquet}, we consider here the special case in which multipliers are real and positive. In more general cases, new function spaces in addition to $\mathcal{P}_T^n$ are required (see Ref.~\cite{Moore:2006kj}). Assuming the multipliers are distinct, the eigenvectors form collectively a basis of $\mathbb{R}^n$. From the multipliers, the Floquet exponents $\lambda_k = \log(\mu_k)/T$ can be calculated, defining the period averaged growth/decay of tangent perturbations initially aligned to the invariant subspaces $\mathbf{e}_k(0)$. 


We introduce the Floquet eigenfunctions $\mathbf{w}_k(t) \in \mathcal{P}_T^n$, generated by the eigenvectors $\mathbf{e}_k(0)$ as
\begin{equation}
	\mathbf{w}_k(t) = \exp\left(-\lambda_k t\right)\mathbf{M}(t, 0)\mathbf{e}_k(0),
\end{equation}
and satisfying the differential eigenvalue problem
\begin{equation}\label{eq:floquet-forward-eigenproblem}
	\mathcal{L} \mathbf{w}_k(t) \equiv \frac{\mathrm{d}\mathbf{w}_k(t)}{\mathrm{d}t} - \mathbf{f}_\mathbf{u}(t)\mathbf{w}_k(t) = - \lambda_k \mathbf{w}_k(t),
\end{equation}
with unit $\|\cdot\|_{\mathcal{P}_T^n}$ norm.
Sorting the Floquet exponents in descending order, we denote by $\chi-1$ the number of positive exponents. It is well known that $\mathbf{w}_\chi(t) =  \mathbf{f}(t)/{\|\mathbf{f}(t)\|_{\mathcal{P}_T^n}}$ is a marginal direction, \textcolor{black}{with exponent $\lambda_\chi = 0$ and multiplier $\mu_\chi=1$}. In other words, the linear differential operator $\mathcal{L}$ is singular, with nullspace
\begin{equation}\label{eq:nullspace-of-direct}
	\mathrm{Null}\left\{ \mathcal{L} \right\} = \mathrm{span}\{{\mathbf{w}_\chi}(t)\}.
\end{equation}

The other useful element of Floquet theory for our purposes, perhaps considered less extensively in the literature \cite{Marcotte:2016hn}, are the adjoint Floquet eigenfunctions $\mathbf{w}^+_k(t)$. These are elements of $\mathcal{P}_T^n$ defined as
\begin{equation}
	\mathbf{w}^+_k(t) = \exp( -\lambda_k(T - t))\mathbf{M}(t, T)^\top \mathbf{e}_k^+(T),
\end{equation}
where the vectors $\mathbf{e}^+_k(T)$ are the left eigenvectors of the monodromy matrix, satisfying
\begin{equation}
	\mathbf{e}^+_k(T)^\top\mathbf{M}(0, T) = \mu_k\mathbf{e}^+_k(T)^\top
\end{equation}
for the same multipliers and exponents of the direct problem. The adjoint eigenfunctions satisfy the adjoint differential eigenproblem
\begin{equation}
	\mathcal{L}^+ \mathbf{w}_k^+(t) \equiv -\frac{\mathrm{d}\mathbf{w}_k^+(t)}{\mathrm{d}t} - \mathbf{f}_\mathbf{u}^\top(t)\mathbf{w}_k^+(t) = -\lambda_k \mathbf{w}^+_k(t),
\end{equation}
where the operator $\mathcal{L}^+$ is the adjoint of $\mathcal{L}$ according to the inner product (\ref{eq:inner-product}). 
These two operators share the same spectrum and the adjoint operator $\mathcal{L}^+$ is thus singular with nullspace
\begin{equation}\label{eq:nullspace-adjoint}
	\mathrm{Null}\left\{ \mathcal{L}^+ \right\} = \mathrm{span}\{{\mathbf{w}^+_\chi}(t)\}.
\end{equation}

A final remark is that, at any time $t$, the direct or adjoint eigenfunctions do not form individually an orthogonal set of vectors but instead satisfy the bi-orthogonality relation 
\begin{equation}\label{eq:to-be-shown-biorthogonality-all-s}
	\mathbf{w}_k^+(t)^\top\cdot\mathbf{w}_j(t) = \delta_{kj}C_k\quad \forall t,
\end{equation}
with $\delta_{kj}$ the Kronecker symbol and with \textcolor{black}{constants} $C_k \neq 0 \in \mathbb{R}$.

\section{Sensitivity analysis}\label{sec:sensitivity-analysis}
We now briefly recall the tangent approach to compute the sensitivity of period averaged quantities \cite{Lasagna:2017tz}. Consider an observable of interest, denoted by a function 
\begin{equation}\label{eq:observable}
J(t) = J(\mathbf{u}(t)) : \mathbb{R}^n \mapsto \mathbb{R},
\end{equation}
that, for the sake of simplicity, is assumed here not to depend explicitly on the parameter $\alpha$. Let also denote the partial derivative of the observable with respect to its argument as $J_\mathbf{u}(t) = J_\mathbf{u}(\mathbf{u}(t))$, mapping $\mathbb{R}^n$ to $\mathbb{R}^n$. Consider one periodic orbit $\mathbf{u}(t)\in\mathcal{P}_T^n$ and define the function $\mathcal{J}(\alpha) : \mathbb{R} \mapsto \mathbb{R}$
\begin{equation}\label{eq:objective-functional}
	\displaystyle \mathcal{J}(\alpha) = \frac{1}{T}\int_0^{T} J(\mathbf{u}(t)) \mathrm{d}t
\end{equation}
as the period average of the observable over the periodic orbit. This is a function of the parameter $\alpha$ since both $\mathbf{u}(t)$ and $T$ depend, implicitly, on it. The goal is to compute the sensitivity of the period average with respect to $\alpha$, the gradient ${\mathrm{d}\mathcal{J}}/{\mathrm{d}\alpha}$ (also shortened to $\mathcal{J}_\alpha$ in the following sections).

Linearisation of (\ref{eq:objective-functional}) as reported in Ref.~[\onlinecite{Lasagna:2018uo}] shows 
that the gradient of the period average is given by the inner product
\begin{equation}\label{eq:gateaux-derivative}
	{\mathrm{d}\mathcal{J}}/{\mathrm{d}\alpha} = \left\llbracket J_\mathbf{u}(t), \mathbf{v}(t)\right\rrbracket,
\end{equation}
where the perturbation $\mathbf{v}(t) \in \mathcal{P}_T^n$ satisfies the tangent equation
\begin{equation}\label{eq:tangent-problem}
	\mathcal{L} \mathbf{v}(t) = \mathbf{f}_\alpha(t) - \tau \mathbf{f}(t).
\end{equation}
The perturbation $\mathbf{v}(t)$ is the first order state-space deformation of the periodic orbit $\mathbf{u}(t)$ when $\alpha$ is varied. The scalar $\tau = (\mathrm{d}T/\mathrm{d}\alpha)/T$ is the (unknown) relative period gradient, producing an algebraically growing mode along $\mathbf{f}(t)$, allowing $\mathbf{v}(t)$ to be time periodic. The introduction of this term is akin to classical approaches in perturbation/continuation analysis of periodic problems \cite{Farkas:2013gm,Viswanath:2001ko}, where time is rescaled by the period. The forcing term $\mathbf{f}_\alpha(t) \in \mathcal{P}^n_T$ is the derivative of the right hand side of (\ref{eq:system}) with respect to the parameter
\begin{equation}
	\mathbf{f}_\alpha(t) = \mathbf{f}_\alpha(\mathbf{u}(t), \alpha) = \frac{\partial \mathbf{f}(\mathbf{u}(t), \alpha)}{\partial \alpha}.
\end{equation}

For \textcolor{black}{a} hyperbolic orbit, the differential operator $\mathcal{L}$ is singular with nullspace given by (\ref{eq:nullspace-of-direct}) and equation (\ref{eq:tangent-problem}) has a one parameter family of solutions. Physically speaking, this is a reflection of the translational invariance along a periodic orbit: if $\mathbf{v}(t)$ is a solution, then $\mathbf{v}(t) + \sigma \mathbf{f}(t)$ is a solution too, for any $\sigma\in\mathbb{R}$, with same gradient $\mathrm{d}\mathcal{J}/\mathrm{d}{\alpha}$, since 
\begin{equation}\label{eq:ju-and-f-are-always-orthogonal}
\llbracket J_\mathbf{u}(t), \mathbf{f}(t) \rrbracket = 0
\end{equation}
for all possible cost functions when $\mathbf{u}(t) \in \mathcal{P}^n_T$. Hence, for (\ref{eq:tangent-problem}) to have a solution, the scalar $\tau$ must have the unique value that shifts the right hand side $\mathbf{f}_\alpha(t) - \tau \mathbf{f}(t)$ in the range of the operator $\mathcal{L}$, or, by Fredholm's Alternative (cf.~[\onlinecite{hale2009ordinary}], Lemma 1.1, pg. 146) makes it orthogonal to the nullspace of its adjoint $\mathcal{L}^+$, i.e.~by satisfying
\begin{equation}
	\left\llbracket\mathbf{w}_\chi^+(t),  \mathbf{f}_\alpha(t) - \tau \mathbf{f}(t)\right\rrbracket = 0.
\end{equation}
Hence, the scalar $\tau$ could be in principle determined as
\begin{equation}\label{eq:period-gradient}
	\tau = \frac{\left\llbracket\mathbf{w}_\chi^+(t), \mathbf{f}_\alpha(t)\right\rrbracket}{C_\chi \|\mathbf{f}(t)\|_{\mathcal{P}^n_T}}
\end{equation}
if $\mathbf{w}_\chi^+(t)$ was known\textcolor{black}{, with $C_\chi$ from (\ref{eq:to-be-shown-biorthogonality-all-s})}. Numerically, it is more convenient to drop the singularity by adding the constraint 
\begin{equation}\label{eq:constraint-tangent}
\left\llbracket \mathbf{v}(t), \mathbf{f}(t)\right\rrbracket = 0,	
\end{equation}
which fixes the component of $\mathbf{v}(t)$ along the nullspace and leads to the solution of (\ref{eq:tangent-problem}) with minimum norm \cite{Moore:2006kj}. In matrix form, the tangent problem reads 
\begin{equation}\label{eq:tangent-problem-block}
	\displaystyle \left[\begin{array}{c|c}
	\displaystyle \mathcal{L} & \mathbf{f}(t) \\
	\hline\!\rule{0in}{.35cm} 
	\left\llbracket \cdot, \mathbf{f}(t) \right\rrbracket               &  0\\
	\end{array} \right]\cdot
	\left[\begin{array}{c}
	\mathbf{v}(t) \\
	\hline\!\rule{0in}{.35cm} \tau
	\end{array}\right] = 
	\left[\begin{array}{c}
	\mathbf{f}_\alpha(t) \\
	\hline\!\rule{0in}{.35cm}0
	\end{array}\right]
\end{equation}
and its solution provides the perturbation $\mathbf{v}(t)$ and the period gradient $\tau$. Note that the left hand side of (\ref{eq:tangent-problem-block}) has the same structure of the Newton-Raphson linear problems arising in the search of periodic solutions \cite{Lan:2004ch,Lasagna:2017tz}, and similar discretization techniques can be employed.

Constraining $\mathbf{v}(t)$ to remain in $\mathcal{P}^n_T$ by using an appropriate numerical method \cite{Lasagna:2017tz} is the key to avoid exponential instabilities intrinsic to the tangent dynamics around an unstable periodic orbit \cite{2000TellA..52..523L}. The solution $\mathbf{v}(t)$ will thus not grow exponentially along the most unstable subspace $\mathbf{w}_1(t)$, but will remain bounded, with magnitude and structure that depend on the complete stability spectrum, as we shall see in section \ref{sec:forward-sensitivity-floquet}. With a bounded $\mathbf{v}(t)$, the gradient (\ref{eq:gateaux-derivative}) is effectively the slope of the function $\mathcal{J}(\alpha)$ obtained from continuation.

\section{Stability and sensitivity}\label{sec:forward-sensitivity-floquet}
In this section, we explain the relation between the linear stability of periodic orbits and the sensitivity of period averages. We do not make use of the following results for our numerical calculations in section \ref{sec:numerical-results}, but aim to develop tools to facilitate their interpretation. 

Fundamentally, the approach consists in projecting the tangent problem (\ref{eq:tangent-problem}) onto the invariant subspaces formed by the Floquet eigenfunctions. With the same assumptions on the multipliers as in section \ref{sec:floquet-analysis}, the solution $\mathbf{v}(t)$ is expanded in the Floquet eigenfunctions,
\begin{equation}\label{eq:v-expansion}
	\mathbf{v}(t) = \sum_{k=1}^n\mathbf{w}_k(t)a_k(t),
\end{equation}
with unknown expansion coefficients $a_k(t) \in \mathcal{P}_T$. The forcing term in (\ref{eq:tangent-problem}) is also similarly expanded
\begin{equation}\label{eq:fp-expansion}
	\mathbf{f}_\alpha(t) = \sum_{k=1}^n\mathbf{w}_k(t)b_k(t), 
\end{equation}
where the functions $b_k(t) \in \mathcal{P}_T$ can be determined by dotting (\ref{eq:fp-expansion}) with the $k$-th adjoint eigenfunction
\begin{equation}\label{eq:bk-coeff}
	b_k(t) = \frac{\mathbf{w}_k^{+}(t)^\top\cdot\mathbf{f}_\alpha(t)}{C_k}, \quad k = 1, \ldots, n,
\end{equation}
where the bi-orthogonality relation (\ref{eq:to-be-shown-biorthogonality-all-s}) is used.


Substituting the expansion (\ref{eq:v-expansion}) into the tangent equation (\ref{eq:tangent-problem}) and using (\ref{eq:floquet-forward-eigenproblem}), produces
\begin{equation}\label{eq:tangent-problem-with-floquet-first-step}
	\sum_{k=1}^n\mathbf{w}_k(t)\left[  \frac{\mathrm{d}a_k(t)}{\mathrm{d}t} - \lambda_k  a_k(t) - b_k(t) +\tau\|\mathbf{f}(t)\|_{\mathcal{P}^n_T} \delta_{k, \chi} \right] = \mathbf{0}.
\end{equation}
Since the  Floquet eigenfunctions form a basis of $\mathbb{R}^n$ for all $t$ by assumption, the term in the square brackets in (\ref{eq:tangent-problem-with-floquet-first-step}) must be zero. We thus obtain a set of decoupled linear ODEs with constant coefficients
\begin{equation}
	 \frac{\mathrm{d}a_k(t)}{\mathrm{d}t} = \lambda_k a_k(t) + b_k(t) -\tau\|\mathbf{f(t)}\|_{\mathcal{P}^n_T} \delta_{k, \chi},
\end{equation}
$k=1\ldots, n$, the tangent sensitivity problem expressed in the basis of the Floquet eigenfunctions. Along the neutral subspace $\mathbf{f}(t)$ the equation reads 
\begin{equation}\label{eq:a_chi_equation}
	\frac{\mathrm{d} a_\chi(t)}{\mathrm{d}t}  = \frac{\mathbf{w}_\chi^+(t)^\top\cdot\mathbf{f}_\alpha(t)}{C_\chi} - \tau\|\mathbf{f}(t)\|_{\mathcal{P}^n_T}.
\end{equation}
In order for $a_\chi(t)$ to remain in $\mathcal{P}_T$, the right hand side must have zero integral over the period by Fredholm's Alternative, since the equation is self adjoint and $a_\chi^+(t) = 1$ is a nontrivial solution of the homogeneous adjoint equation. This constraints fixes the relative period gradient $\tau$ to a value that is the same as equation (\ref{eq:period-gradient}).

A particularly insightful expression can be derived for the expanding and contracting directions. The solution of the scalar ODEs ${\mathrm{d}a_k(t)}/{\mathrm{d}t} = \lambda_k a_k(t) + b_k(t)$ can be expressed with a Green's function approach (cf.~[\onlinecite{hale2009ordinary}], pg.~148) as
\begin{equation}
	a_k(t) = \int_0^{T} \frac{\mu_k}{1 - \mu_k}e^{-\lambda_k s}b_k(t + s)\,\mathrm{d}s,
\end{equation}
with \textcolor{black}{$\mu_k$ the Floquet multipliers}, and the upper bounds
\begin{align}\label{eq:upper-bound-on-solution}
	\sup_{t} |a_k(t)| \le\;& \sup_{t} |b_k(t)|/{|\lambda_k|}  = B_k/{|\lambda_k|},
\end{align}
the key result of this section, can be derived. 

This bound suggests several remarks. First, without further details on the coefficients $b_k(t)$, generic parameter perturbations induce relatively small state-space changes along the highly contracting or expanding directions, while most of the ``yield'' occurs along the Floquet invariant subspace associated to Floquet exponents with small magnitude. This is in stark contrast with classical sensitivity analysis methods for chaotic trajectories \cite{2000TellA..52..523L,Eyink:2004gk}, where only the most unstable covariant Lyapunov vector \cite{Ginelli:2007ja,Wolfe:2007jq} dominates asymptotically the solution of the tangent equations \cite{2000TellA..52..523L}. When varying the system parameter $\alpha$ towards a bifurcation, one (or a pair of) Floquet exponent crosses the imaginary axis and the tangent solution displays a large amplitude along the corresponding direction, resulting in large gradients of time averaged quantities. 

Second, the bound (\ref{eq:upper-bound-on-solution}) shows that the amplitude of the tangent solution along a particular Floquet eigenfunction $\mathbf{w}_k(t)$ depends directly on the strength of the projection of the forcing $\mathbf{f}_\alpha(t)$ on the associated adjoint Floquet eigenfunctions, the coefficients $b_k$. Hence, for spatially extended systems, knowledge of the spatiotemporal dynamics of the adjoint Floquet eigenfunctions and not just the direct ones \cite{Xu:2018gf,2018JFM...849..942N,Inubushi:2015ja}, might provide an understanding of how physically relevant features of the solutions are influenced by problem parameters \cite{Marcotte:2016hn}. One can then interpret the adjoint Floquet eigenfunctions as special directions where the forcing $\mathbf{f}_\alpha(t)$ can be particularly effective in modifying dynamical behaviour \cite{Luchini:2014fv}, which is useful, for instance, for control design \cite{Giannetti:2019do}.

Third, and most importantly, the boundedness of the forcing term $\mathbf{f}_\alpha(t)$ and of the Floquet eigenfunctions implies that the coefficients $b_k(t)$, and thus the expansion coefficients $a_k(t)$ and the tangent solution $\mathbf{v}(t)$, have, on average, similar magnitude for long periodic orbits if the exponents of long periodic orbits converge as $T\rightarrow\infty$. At this stage it is convenient to note that the Floquet exponents are the period averages of the ``local exponents'' \cite{Bosetti:2014cv,Ding:2016iw}
\begin{equation}\label{eq:local-floquet-exponent}
	\lambda_k(t) = \frac{\mathbf{w}_k(t)^\top [\mathbf{f}_\mathbf{u}^\top(t) + \mathbf{f}_\mathbf{u}(t)]\mathbf{w}_k(t)}{\|\mathbf{w}_k(t)\|^2},
\end{equation}
uniquely defined functions of state space \cite{Trevisan:1998km} expressing the local growth rate of tangent perturbations along the invariant subspaces (here $\|\cdot\|$ indicates the \textcolor{black}{Euclidean} norm).
By the Central Limit Theorem, the distribution of the $k$-th Floquet exponent across distinct orbits of similar period $T$, must converge in law to a Dirac delta function \cite{Ott:2002wz} with standard deviation decaying as $T^{-1/2}$, assuming that the auto-correlation of time histories of (\ref{eq:local-floquet-exponent}) decays sufficiently quickly \cite{Ott:2002wz}. 
Hence, the bound (\ref{eq:upper-bound-on-solution}) indicates that the distribution of the sensitivity of period averages $\mathrm{d}\mathcal{J}/\mathrm{d}\alpha$ will also converge to a delta function as $T$ increases. In other words, while some scatter might be observed for short cycles, long periodic orbits will asymptotically provide the same sensitivity to parameter perturbations.

\section{Numerical results}\label{sec:numerical-results}
To answer the question posed in the introduction, we now turn to numerical experiments and consider periodic orbits of two well-known chaotic systems. The first is given by the Lorenz equations \cite{Lorenz:1963tf,Sparrow:2012hi}
\begin{equation}
  	\begin{cases}\label{eq:lorenz-equations}
	\mathrm{d}u_1/\mathrm{d}t &=\, \sigma(u_2-u_1),\\
	\mathrm{d}u_2/\mathrm{d}t &=\, \rho u_1 - u_2 - u_1u_3,\\
	\mathrm{d}u_3/\mathrm{d}t &=\, u_1u_2  - \beta u_3,
  \end{cases}
\end{equation}
where standard parameters $\sigma=10$, $\beta=8/3$ and $\rho = 28$ are used throughout. 
As in other sensitivity studies on the Lorenz equations \citep{2000TellA..52..523L,Eyink:2004gk, Wang:2013cx, Liao:2016hn, Lasagna:2018uo,Craske:2019fs}, we consider the sensitivity of the period average of the observable $J(t) = u_3(t)$ with respect to perturbations of $\rho$. Numerical integration of chaotic trajectories is performed using a classical fourth-order Runge-Kutta method with $\Delta t = 0.005$.


The second system is a finite-dimensional truncation of a spatially extended system, the Kuramoto-Sivashinky equation
\begin{equation}\label{eq:ks-equation}
	\frac{\partial u}{\partial t} + u\frac{\partial u}{\partial x} + \frac{\partial^2 u}{\partial x^2} + \nu \frac{\partial^4 u}{\partial x^4} = 0,
\end{equation}
defining the evolution of zero-mean, spatially-periodic fields $u \equiv u(x, t)$ over the domain $x\in[0, 2\pi]$ with dynamics restricted to the invariant subspace of odd solutions \cite{Lan:2008kg}. Here, we consider a relatively high diffusivity constant $\nu = (2\pi/L)^2$, with $L=39$, the same value we considered in previous work \cite{Lasagna:2017tz}. The spectral expansion
\begin{equation}
	\displaystyle u(x, t) = \sum_{k=-n}^n i u_k(t) \exp(i k x),
\end{equation}
with $u_k = -u_{-k}$, $u_0 = 0$, is truncated at $n=28$, leading to a system of ODEs approximating solutions of the original partial differential equation.
Some of the numerical results reported in the next sections have been checked in a statistical sense at finer resolutions, with negligible quantitative changes. 
We take the energy density 
\begin{equation}\label{eq:energy-density}
	J[u(x, t)] = \frac{1}{4\pi}\int_0^{2\pi} u^2(x, t)\,\mathrm{d}x
\end{equation}
as the functional of interest and examine the sensitivity of its average with respect to the diffusivity $\nu$. Numerical integration of chaotic trajectories is performed using the fourth-order accurate implicit-explicit method \texttt{IMEXRKCB4} \cite{Cavaglieri:2015gr}, with time step $\Delta t = 0.125\nu$.

\subsection{Search of long periodic orbits}
We use a global Newton-Raphson search algorithm developed in previous work \cite{Lasagna:2017tz}, based on the original method of Ref.~[\onlinecite{Lan:2004ch}] and classical techniques for nonlinear boundary value problems \cite{Ascher:1994ty}. Briefly, at iteration $k$, this method solves a Newton-Raphson update equation to adjust a trial solution composed of a state-space loop $\mathbf{u}_k\in\mathcal{P}^n_{T_k}$ and a period $T_k$. The loop is not, at least initially, a solution of the equations, i.e.~the residual
\begin{equation}\label{eq:residual-search}
	\mathbf{r}_k(t) = \mathrm{d}{\mathbf{u}}_k(t)/\mathrm{d}t - \mathbf{f}(\mathbf{u}_k(t), \alpha) \; \in \mathcal{P}_{T_k}^n
\end{equation}
is generally different from zero along the loop. The only significant modification that we have implemented is that the loop derivative operator $\mathrm{d}/\mathrm{d}t$ is approximated using an eight-order accurate finite-difference stencil (instead of fourth order), enhancing the overall accuracy/cost ratio and allowing longer orbits to be found. The same high-order discretisation is used for the solution of the tangent problem (\ref{eq:tangent-problem-block}). Initial guesses are obtained from near recurrences of the chaotic flow.

In previous work \cite{Lasagna:2017tz}, we attempted to locate exhaustively all short periodic orbits and examined their sensitivity as a function of the topological length. In this work, we adopt a different strategy, motivated by the objective of examining the properties of \textit{typical} long periodic orbits found by Newton-Raphson searches. To this end, rather than considering the topological length, we select a number of arbitrary reference periods, denoted as $\hat{T}$, and locate up to five thousand periodic orbits with actual period falling within $\pm 5$\% of the reference period. Periodic orbits do have inherent time scales (the period of the shortest cycle), but for long reference periods we have observed that the actual period of converged solutions is uniformly distributed in the $\pm 5$\% range. Hence, the reference periods can be selected arbitrarily and are chosen here such that the maximum reference period $\hat{T}_{\mathrm{max}}$ is about two orders of magnitude larger than the period of the shortest admissible cycle, as indicated in table \ref{tab:details}. This range is sufficiently wide to reveal the asymptotic convergence of properties of long periodic orbits as the period increases. 

\begin{figure}[htbp]
	\centering
	\includegraphics[width=0.49\textwidth]{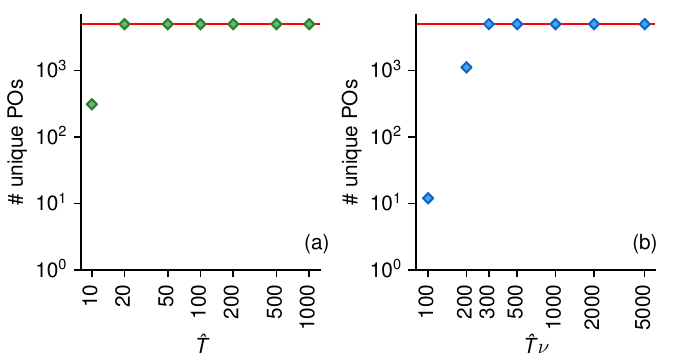}
	\caption{Number of unique periodic orbits found as a function of the reference period $\hat{T}$, for the Lorenz equations, panel (a), and KS system, panel (b). The red line is at 5000. For the KS system, the reference period is scaled by the diffusivity.}
	\label{fig:search-results}  
\end{figure}

Search results are reported in figure \ref{fig:search-results}. 
For short reference periods, we locate as many orbits as it is feasible and the number of periodic orbits found grows exponentially with the period. The number of periodic orbits found quickly saturates the set threshold of five thousand orbits. We stress the fact that our focus is not to provide a description of the statistical distribution of the \textit{complete} set of periodic orbits of high topological lengths, nor to use such quantities to approximate the measure using cycle averaging theory. Such calculations would be biased by the sampling. Rather, we aim to develop an understanding of what to expect from long orbits that might be \textit{typically} found numerically by converging near recurrence events. Hence, the threshold is chosen so that statistics over the ensemble are sufficiently robust.

\begin{figure}[htbp]
	\centering
	\includegraphics[width=0.49\textwidth]{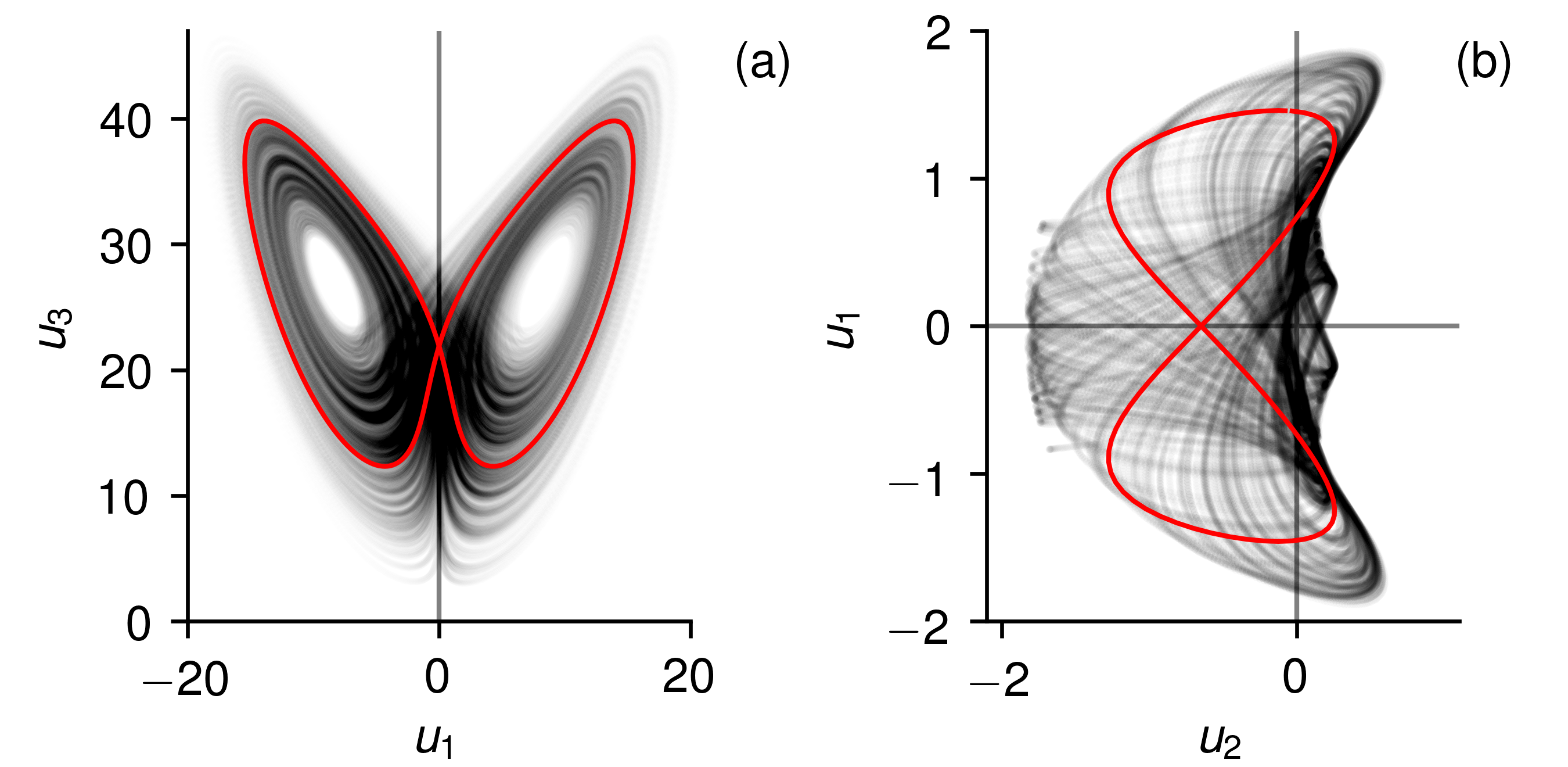}
	\caption{Projections of the shortest (solid red) and longest (thin black) periodic orbits found in this study for the Lorenz, panel (a), and KS equations, panel (b).}
	\label{fig:short-long}
\end{figure}

We report in figure \ref{fig:short-long} the shortest and longest periodic orbits found for the Lorenz equations, panel (a), and KS system, panel (b). While the short cycles are topologically simple, longer orbits wind around the attractor in a complicated fashion and are thus indistinguishable to the eye from long chaotic trajectories. 
\begin{table}
\caption{\label{tab:details}Reference temporal grid spacing for the finite-difference approximation of the derivative involved in the search of periodic orbits, period of the shortest admissible orbit and ratio between the largest reference period and $T_\mathrm{min}$.}
\begin{ruledtabular}
\begin{tabular}{lllll}
 &reference $\Delta t$ & $T_\mathrm{min}$ & $\hat{T}_{\mathrm{max}}/T_\mathrm{min}$\\
\hline
Lorenz& 0.01 & 1.5586 & $\sim$ 645\\
KS& 0.125$\nu$ & 24.9080$\nu$ & $\sim$ 201
\end{tabular}
\end{ruledtabular} 
\end{table}
\begin{figure}[htbp]
	\centering
	\includegraphics[width=0.49\textwidth]{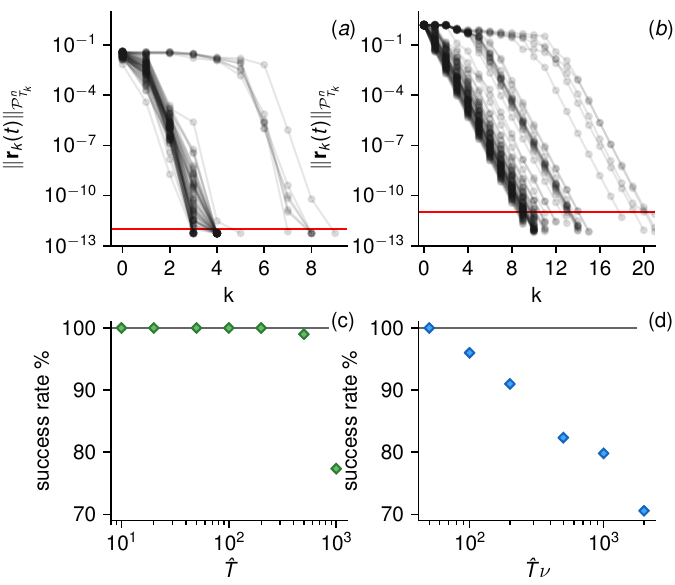}
	\caption{Convergence of the norm of the residual (\ref{eq:residual-search}) during the Newton-Raphson search of long periodic solutions of the Lorenz equation, panel (a), and KS equations, panel (b), for one hundred test searches with reference period $\hat{T}=1000$, and $\hat{T}=5000\nu$, respectively. Only converged searches are reported. The horizontal red line indicates the stopping tolerance. Panels (c) and (d): success rate of the search of periodic orbits estimated from the same test searches, as a function of the reference period.}
	\label{fig:NR-plot} 
\end{figure}
We build an inventory of unique solutions by ensuring that all periodic orbits found have period average (\ref{eq:objective-functional}) differing to at least eight decimal places. This is few orders of magnitude larger than the accuracy at which this quantity is determined in the search, with the temporal discretisation settings reported in table \ref{tab:details}. Because of the high number of orbits existing at such high topological lengths, very few duplicates have been found.
For both systems, we have observed that relatively few Newton-Raphson iterations are required and that the convergence history is independent of the period $T$. However, the success rate appears to decline slightly for longer periods, arguably as a result of the increasing condition number of the Newton-Raphson update problem. This is illustrated in figure \ref{fig:NR-plot}, where we show the history of the norm of the residual (\ref{eq:residual-search}) for about a hundred long period searches for the Lorenz and KS equations, in panels (a) and (b), respectively. Panels (c) and (d) shows how the success rate, estimated from these test searches, varies with the period. We have observed these trends to be independent of the temporal discretisation and, for the KS equations, of the spatial resolution. This gives us enough confidence that these orbits are numerically reliable approximations of exact solutions of the equations and not an artefact of the search method.

In the next sections, we examine properties of these orbits. We first focus on period averages and Floquet exponents in sections \ref{sec:period-averages} and \ref{sec:floquet-exponents}, respectively. To characterise how properties vary across the ensemble of orbits at each reference period, we compute the ensemble mean and standard deviation and denote these quantities by $\mathrm{mean}[\cdot]$ and $\mathrm{std}[\cdot]$, respectively. We do not consider these moments as substitutes of cycle averaging formulae but we use them to characterise in statistical terms the properties of typical orbits obtained numerically. Probability distributions of these quantities are also shown, with the caveat that they only represent \textit{typical} orbits, and not the complete set of admissible periodic orbits. Sensitivity of period averages are then finally considered in section \ref{sec:sensitivity-of-period-averages}. We also compare averages, exponents and sensitivities to analogous quantities computed on chaotic trajectories, to address the original question whether long periodic orbits can be considered as accurate proxies for the chaotic state.

\subsection{Statistics of time averages}\label{sec:period-averages}
\begin{figure}[htbp] 
	\centering 
	\includegraphics[width=0.49\textwidth]{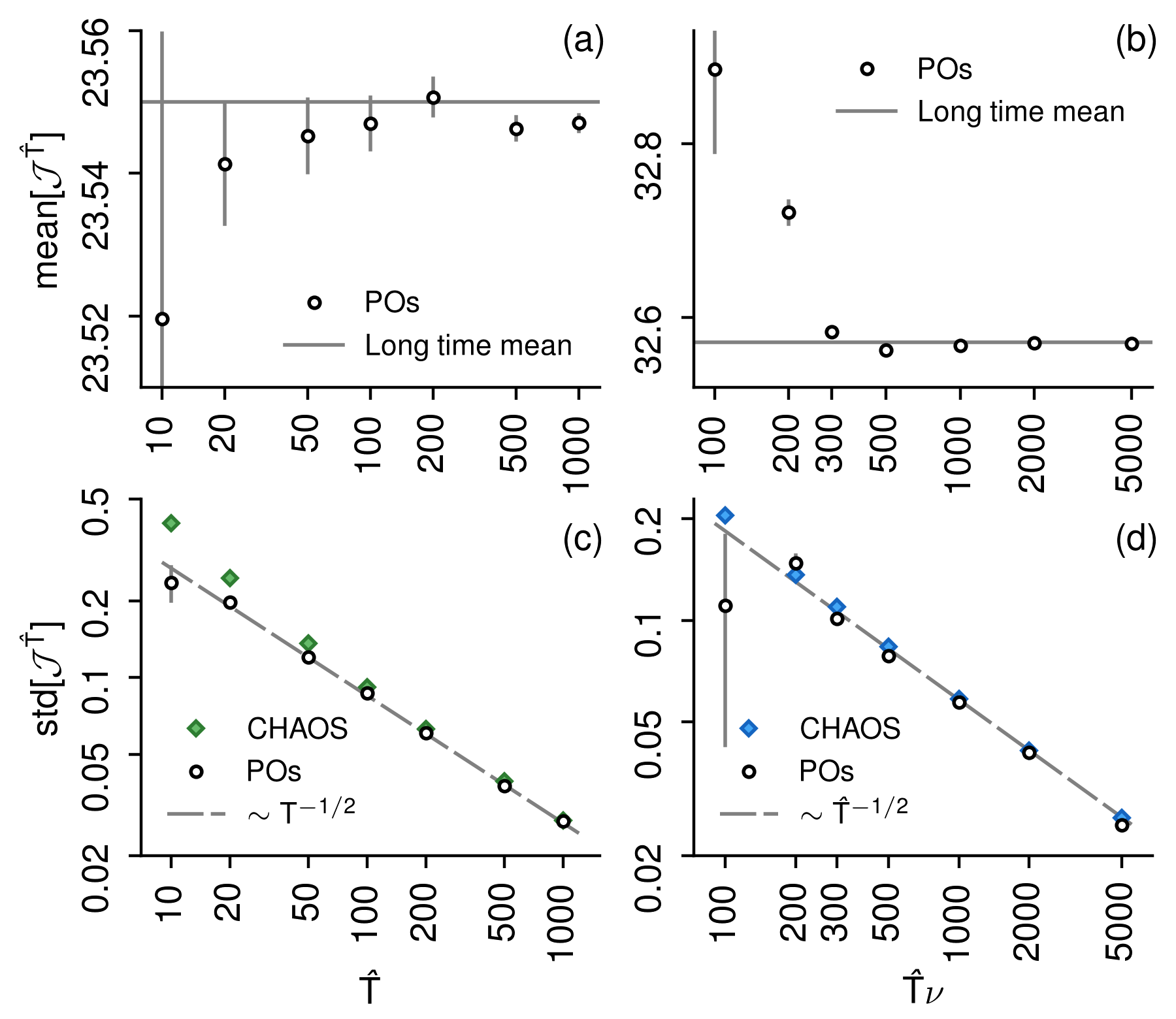} 
	\caption{Ensemble mean, panels (a, b), and standard deviation, panels (c, d), of the time averaged observable as a function of the reference period $\hat{T}$, for chaotic trajectories (filled diamonds) and periodic orbits (open circles). In Left panels, Lorenz equations, right panels, KS system.}
	\label{fig:std-mean} 
\end{figure}

The statistics of period averages over typical long periodic orbits are compared to statistics of time averages of chaotic trajectories of same reference length $\hat{T}$ in figure \ref{fig:std-mean}. Here and in subsequent figures, error bars denote plus/minus three times the standard error \cite{Rao:2009tq}.
The notation $\mathcal{J}^{\hat{T}}$ emphasizes the dependence of averages on the reference period. Panels (a) and (b) show the mean period average of typical periodic orbits, for the Lorenz and KS systems, compared to the long time average from chaotic simulations. \textcolor{black}{For chaotic trajectories the time average does not depend on $\hat{T}$, since the average is a linear operation, and the ensemble mean of short-time averages coincides with the long-time mean.}
For periodic orbits, we observe a small positive/negative distortion at low periods for the KS/Lorenz equations, in agreement with previous work \cite{Zaks:2010ex}. This distortion decays as $\hat{T}$ increases. More importantly, in statistical terms, the period average of typical long periodic orbits spanning increasingly larger fractions of the attractor \textcolor{black}{appears to converge, with the limitations of the present setup, to the long-time average of chaotic trajectories. The asymptotic difference between predictions from periodic orbits and chaotic trajectories is small in relative terms (about 0.0126\%) and may be attributed to the finite observation time ($\hat{T}=1000$) used for periodic orbits. }
 
The standard deviation of the finite-time average over chaotic trajectories, panels (c) and (d), decays asymptotically as $\hat{T}^{-1/2}$. This is the trend predicted for an ensemble of averages by the Central Limit Theorem \cite{Ott:2002wz} if correlations decay sufficiently fast. In fact, both systems considered here display `typical' chaos (in the terminology of Ref.~[\onlinecite{Prasad:1999jt}]), with correlations dying out exponentially. Asymptotically, the standard deviation of the period average decays in the same manner, i.e. longer periodic orbits provide more accurate descriptions of the long-time mean. On the other hand, for short periods, the standard deviation of chaotic trajectories decays at a faster rate for both systems considered. We argue that this is an effect of correlations affecting the asymptotic behaviour. For short periodic orbits, increasing $\hat{T}$ does not necessarily result in a lower variance, in agreement with previous observations \cite{Saiki:2009hm,Zaks:2010ex}. \textcolor{black}{In addition, the standard deviation of period averaged quantities over the ensemble of periodic orbits is lower than that of chaotic trajectories. This is likely a results of the fact that chaotic trajectories are permitted to visit low-probability regions of the attractor with extreme values of the function under average, while periodic orbits with short period are highly constrained and their exploration of the attractor is less pronounced. } Overall, the data in figure \ref{fig:std-mean} suggests that typical long periodic orbits found in computations have, in statistical terms, similar properties to those of chaotic trajectories. 

\begin{figure}[htbp]
	\centering
	\includegraphics[width=0.49\textwidth]{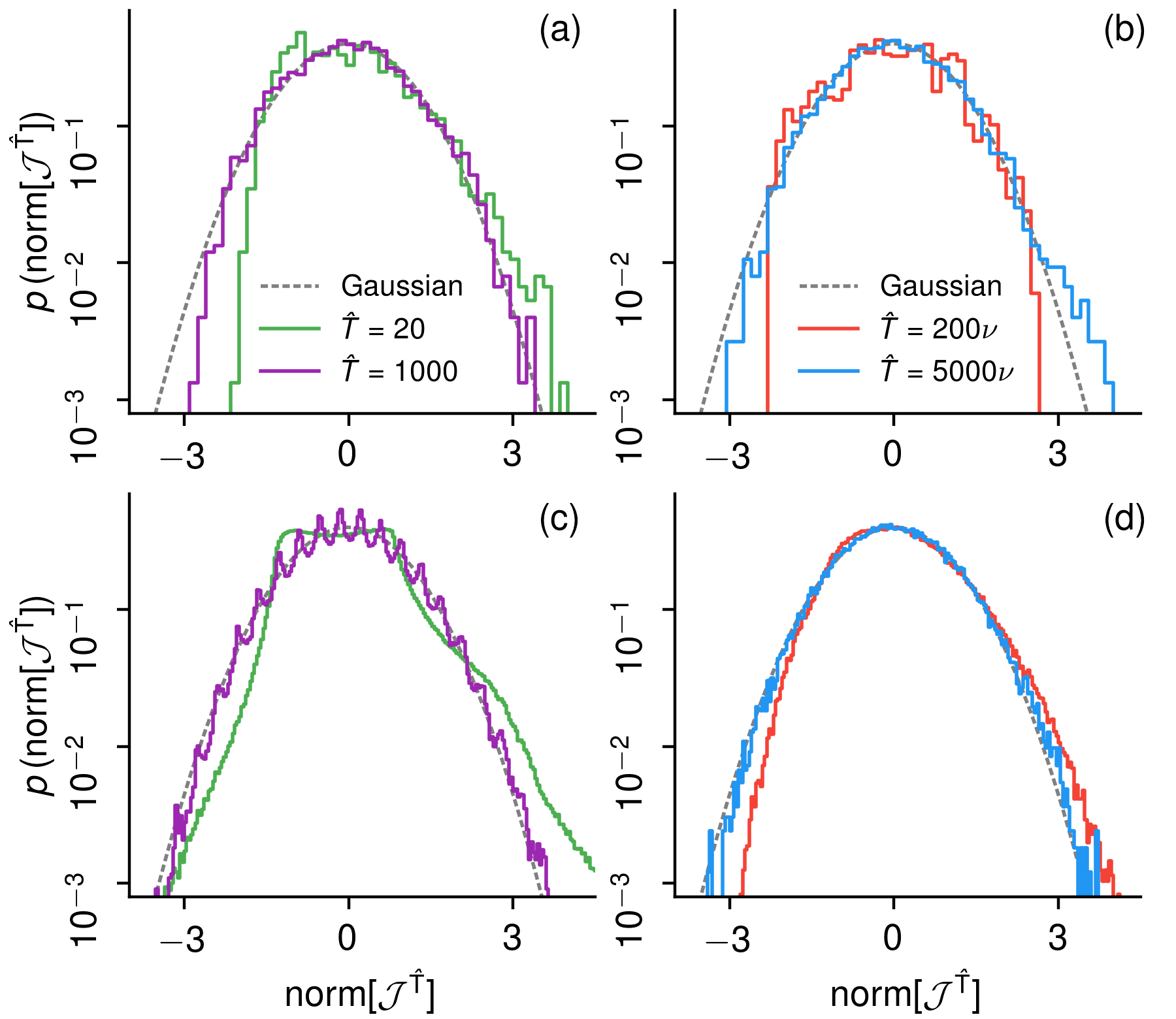}
	\caption{Probability distributions of time averages over periodic trajectories, panels (a) and (b), and chaotic trajectories, panels (c) and (d), for two different reference time spans $\hat{T}$. Left panels, Lorenz equations, right panels, KS system.}
	\label{fig:pdfs-means} 
\end{figure}  

Probability distributions of time averages over periodic and chaotic trajectories are reported in figure \ref{fig:pdfs-means} for two different $\hat{T}$. Data for periodic and chaotic trajectories is reported in panels (a-b) and (c-d), respectively. Data for the Lorenz equations and the KS system is reported in the left and right panels, respectively. Before computing the distributions, samples are normalized such as to have zero mean and unit variance, and the empirical distributions are compared to the normal distribution, indicated in grey in the figure. The distributions are skewed to positive values for short reference time spans. For the Lorenz equations this was observed in previous work \cite{Saiki:2009hm}. \textcolor{black}{The probability density function in figure \ref{fig:pdfs-means}-(c) for $\hat{T}=1000$ is characterised by many small peaks. These can be attributed to the oscillatory nature of solutions of the Lorenz equations, producing an interaction between the oscillation period and the averaging time $\hat{T}$. In fact, the width and spacing of these peaks decreases with the period $\hat{T}$}. Eventually, however, the distributions of averages over chaotic and periodic trajectories appear to collapse to the normal distribution, although less pronouncedly for periodic orbits. Similar to the trends of the standard deviations in figure \ref{fig:std-mean}, this is the behaviour dictated by the Central Limit Theorem \cite{Ott:2002wz}, for which the probability distribution of time averaged quantities can asymptotically be approximated near its peak and within few standard deviations by a Gaussian probability distribution.  These results support the observations reported in Ref.~\cite{Zaks:2010ex}, in which the distributions of period averages converge to delta functions as $\hat{T} \rightarrow \infty$. 
 
\subsection{Statistics of Floquet exponents}\label{sec:floquet-exponents}
We now compare statistics of Floquet exponents of typical periodic orbits with finite-time Lyapunov exponents (FTLE) calculated over chaotic trajectories with same reference period. We calculate the FTLEs using classical methods \cite{Benettin:1980ek,1979PThPh..61.1605S}, involving propagating a set of vectors in tangent space and occasionally performing a re-orthogonalization using the \textcolor{black}{Gram-Schmidt} procedure to counter the inevitable alignment to the most unstable subspace. 
 \begin{figure}[htbp]
	\centering
	\includegraphics[width=0.49\textwidth]{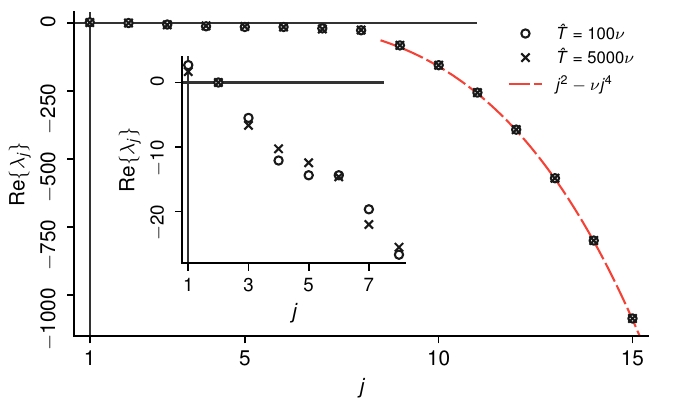}
	\caption{The real part of the fifteen leading Floquet exponents of a long and a short orbit of the KS system. The inset shows the eight leading exponents. The ninth exponent has a much lower value, at around -82.} 
	\label{fig:floquet-single}
\end{figure}
\begin{figure*}[htbp]
	\centering
	\includegraphics[width=0.99\textwidth]{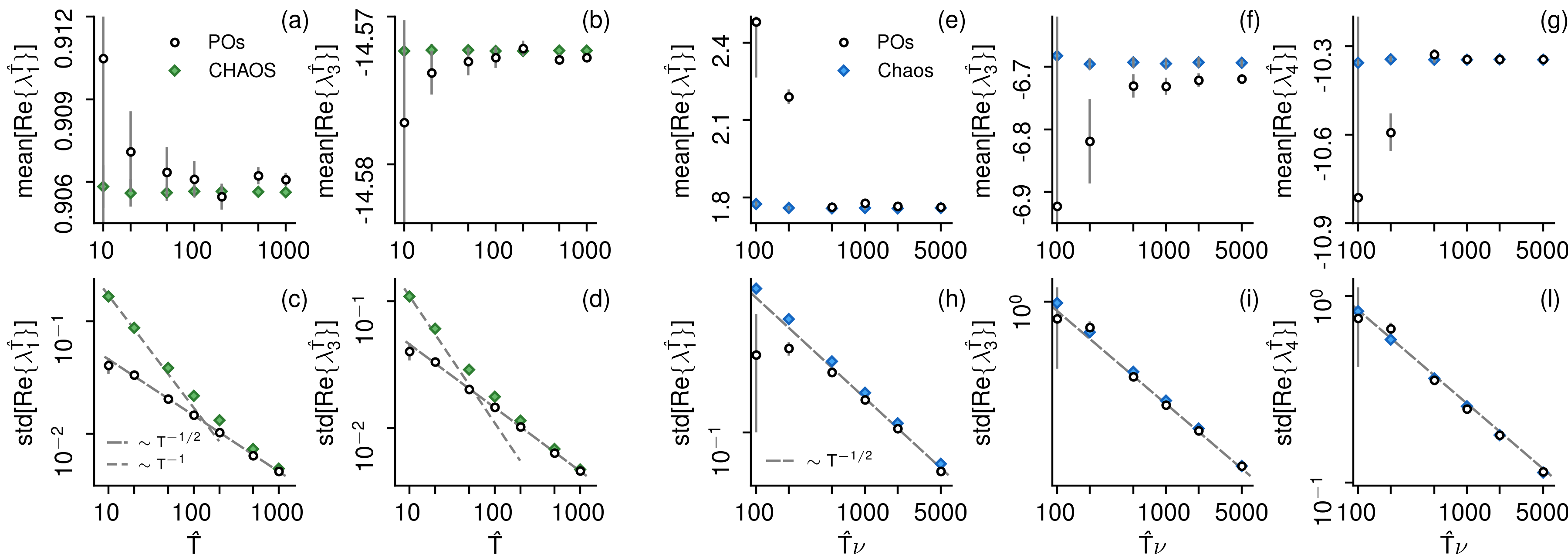}
	\caption{Evolution of mean (top panels) and standard deviation (bottom panels) of few selected Floquet (open circles) exponent and FTLEs (filled diamonds) as a function of the reference period for the Lorenz equations, panels (a) to (d), and for KS system, panels (e) to (l).}
	\label{fig:std-mean-floquet}
\end{figure*}
Computing the spectrum of Floquet exponents is, however, a notoriously challenging problem \cite{Lust:2001iy,Ding:2014vl}, even for orbits of moderate period. The exponential growth of the entries of the monodromy matrix (\ref{eq:eigenvalues-of-monodromy}) makes its eigenvalue decomposition inaccurate in finite-precision arithmetic. In this work, where the focus is on long periodic solutions, we have used a more robust algorithm recently discussed in Ref.~[\onlinecite{Ding:2014vl}]. With this algorithm we have been able to compute accurately Floquet exponents for arbitrarily long orbits, corresponding to multipliers spanning thousands of orders of magnitude. Our implementation does not make use of iterative QR-based eigenvalue algorithms, but works in a matrix-free fashion and only requires an existing time-stepper code for the linearised equations. The algorithm and a small addition to Ref.~\cite{Ding:2014vl} are presented in appendix \ref{app:numerical-method-long-periodic orbits-floquet} for completeness.

For the KS system, all periodic orbits found have only one unstable Floquet eigendirection, and with the present spatial resolution twenty-six contracting directions. For illustrative purposes, the leading part of the Floquet spectrum of one long and one short periodic orbit is shown in figure \ref{fig:floquet-single}. The first eight exponents fall in the range $(-30, 5)$, while the ninth exponent is sharply more negative and is followed by a long tail of negative exponents. These correspond to contracting ``spurious'' modes \cite{Ding:2016iw,Yang:2009jja}, with a value that is closely determined by the linear term of the governing equations. All orbits in our database have a similar spectrum.

We report in figure \ref{fig:std-mean-floquet} data for the two non-trivial exponents of the Lorenz equations, panels (a) to (d) and the first three non-trivial exponents for the KS system, panels (e) to (l). The evolution of the mean (top five panels) and standard deviation (bottom five panels) of selected Floquet exponents (open circles) and FTLEs (filled diamonds) is reported as a function of the reference period $\hat{T}$. Similar to figure \ref{fig:std-mean}, error bars define plus/minus three times the standard error. These are shown only for Floquet exponents, since statistics of the FTLEs are computed over a sufficiently large collection of independent orbits to make the bars smaller than the symbols in the figure. As the reference period increases, the average Floquet exponents of typical orbits \textcolor{black}{generally} converge, within the statistical relevance of our ensemble, to the corresponding infinite-time Lyapunov exponents. In other words, typical long periodic orbits found using Newton-Raphson searches have the same stability properties of long ergodic trajectories. \textcolor{black}{Similar to period averages in figure \ref{fig:std-mean}, a small bias between long periodic orbits and long chaotic trajectories can be observed, likely due to the finite reference periods $\hat{T}$ used in these calculations. In addition, we also} observe a distortion over short periodic trajectories. The standard deviation of exponents of periodic and chaotic trajectories decays \textcolor{black}{asymptotically} as $\hat{T}^{-1/2}$. For short periods, the standard deviation of FTLEs of the Lorenz equations decreases more rapidly, as $\hat{T}^{-1}$. This is induced by exponential tails characterizing the distribution of short-time FTLEs, often observed for intermittent systems \cite{Prasad:1999jt}. \textcolor{black}{On the other hand, the standard deviation of Floquet exponents for short orbits can be lower than asymptotic trends, for the same mechanism outlined for period averages in section \ref{sec:period-averages}.}
\begin{figure*}[htbp]
	\centering
	\includegraphics[width=1.00\textwidth]{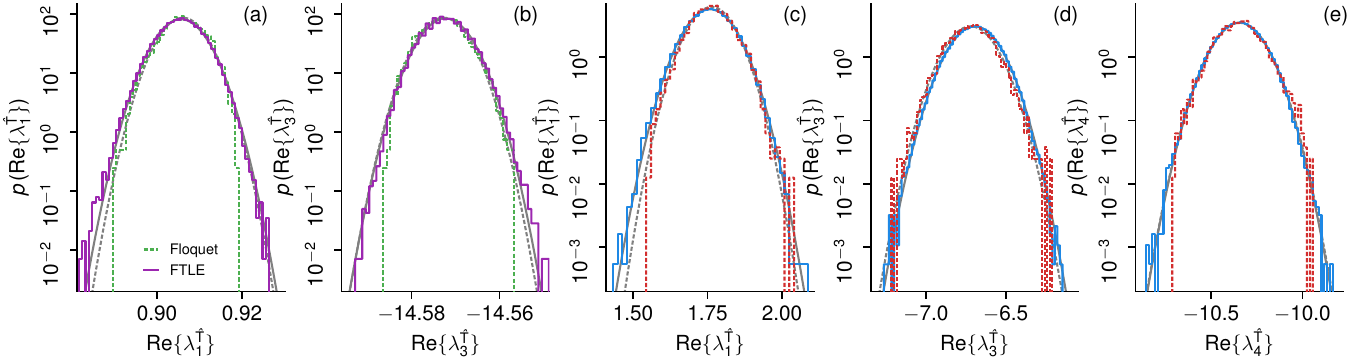}
	\caption{Probability distributions of the two non-trivial Floquet exponents and FTLE for the Lorenz equations at $\hat{T}=1000$, panels (a) and (b), and the first three non-trivial exponents of the KS system at $\hat{T} = 5000$, panels (c) to (e). The parabolas denote Gaussian distributions with mean and standard deviation equal to the sample mean and variance of the numerical data.} 
	\label{fig:pdfs-floquet} 
\end{figure*} 

 \begin{figure}[htbp]
	\centering
	\includegraphics[width=0.49\textwidth]{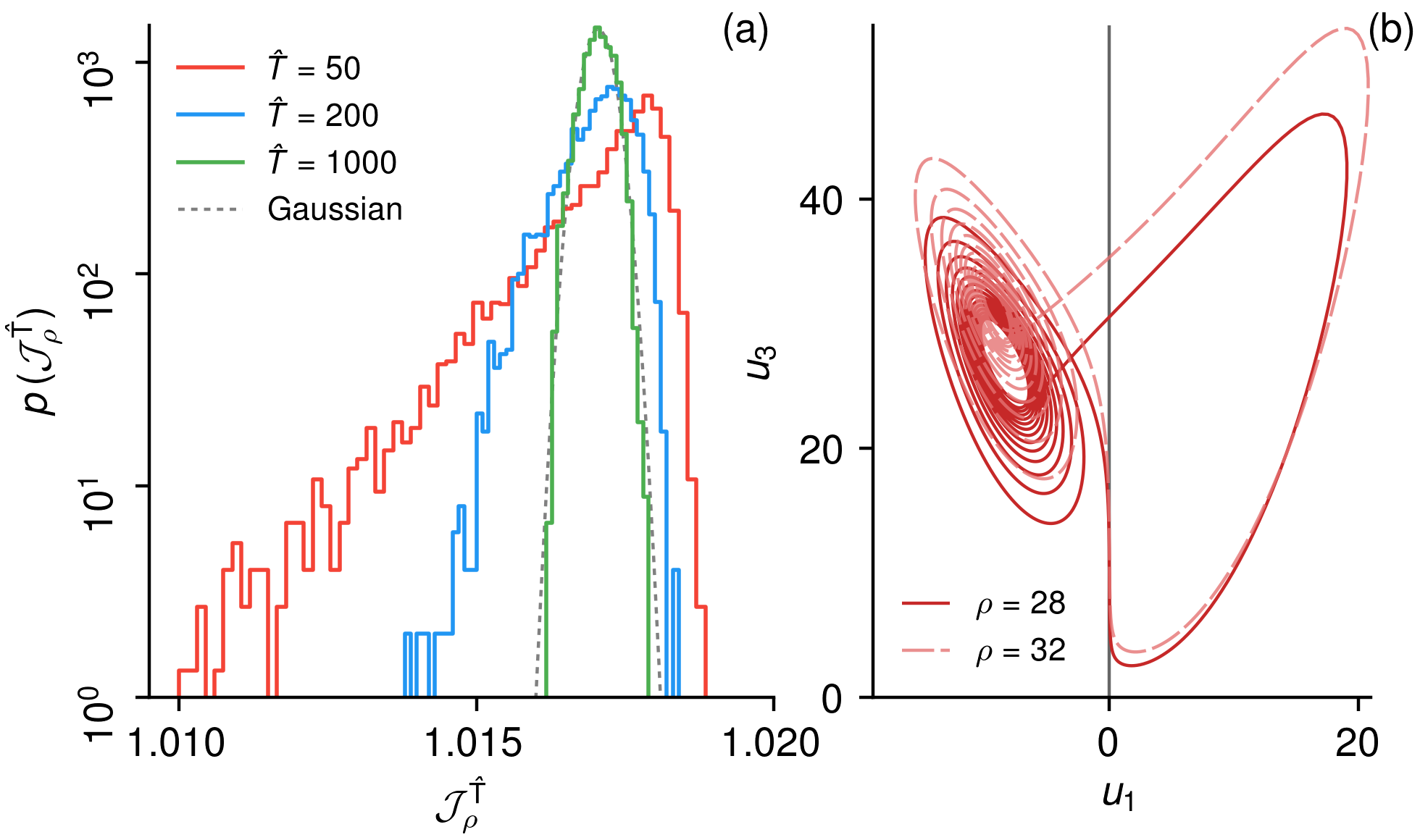}
	\caption{Panel (a): probability distributions of the gradient $\mathcal{J}_\rho^{\hat{T}}$ for typical long periodic orbits of the Lorenz equations. Panel (b): $(u_1, u_3)$ projection of the orbit $\mathrm{A}^{14}\mathrm{B}$ with period $T=10.4965...$ and extremal gradient $\mathcal{J}_\rho = 0.9988...$, for two values of $\rho$.}
	\label{fig:pdfs-sens-lorenz}
\end{figure}

Probability distributions of the exponents are reported in figure \ref{fig:pdfs-floquet}, where we present data for the longest reference period considered to illustrate the asymptotic behaviour. Normal distributions with mean and variance equal to the sample mean and variance of the numerical data are reported in grey. Since Floquet exponents and FTLEs are averages of local quantities (\ref{eq:local-floquet-exponent}), their distributions follow the same trend as that of the quantity $\mathcal{J}^{\hat{T}}$. Hence, for asymptotically long periods, the distributions collapse to the same normal behaviour, at least within plus/minus five standard deviations shown in the figure. 



In summary, the distributions of the Floquet exponents of typical periodic orbits found in our computations localise around the Lyapunov exponents of the chaotic flow. Based on the discussion of section \ref{sec:forward-sensitivity-floquet} and the bound (\ref{eq:upper-bound-on-solution}), this suggests that the distribution of the sensitivity of period averages of typical long orbits will also localise around an asymptotic value. This localisation is examined in the next section.


\subsection{Statistics of sensitivities}\label{sec:sensitivity-of-period-averages}
The probability distribution of gradients $\mathcal{J}_\rho^{\hat{T}}$ from periodic orbits of the Lorenz equations is reported in figure \ref{fig:pdfs-sens-lorenz}-(a), for three reference periods. For short periods, the probability distributions are not normal, but have a heavier left tail that decays fast enough for the first and second central moments to be finite. Extremal periodic orbits with low sensitivity in the left tails feature close passes to the unstable equilibrium at the origin \cite{Eckhardt:1994dv}. Continuation in $\rho$ of the extremal orbit found for $\hat{T} = 10$ (with symbol sequence $\mathrm{A}^{14}\mathrm{B}$ in the notation of Ref.~[\onlinecite{Viswanath:2003gb}]), reported in panel (b), shows that points near the origin of state space move towards the origin, thus causing a lower sensitivity of the $u_3$ variable. For the longer orbits found in our computations from near recurrences, we observe that the fraction of the period spent in the neighbourhood of the origin diminishes in relative terms and approaches that of long chaotic trajectories. As a result, the left tail of the distributions in figure \ref{fig:pdfs-sens-lorenz}-(a) shows a progressively faster decay as $\hat{T}$ increases and the distribution ultimately converges to a Gaussian law (denoted with a dashed line), localised around $\mathcal{J}_\rho^{\hat{T}} \simeq 1.017$.

 \begin{figure}[htbp]
	\centering 
	\includegraphics[width=0.49\textwidth]{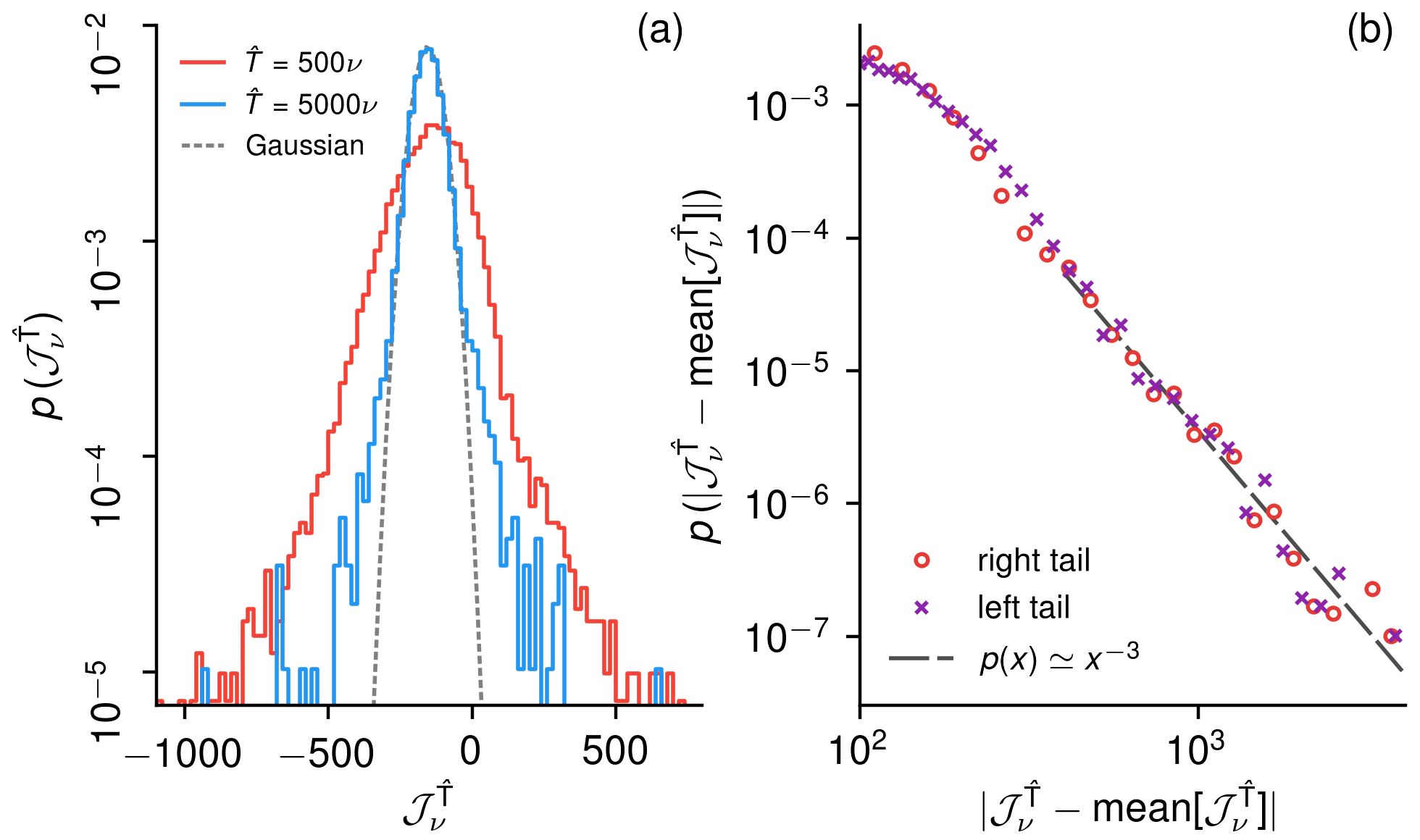}
	\caption{Panel (a): probability distributions of the gradient $\mathcal{J}_\nu^{\hat{T}}$ for periodic orbits of the KS system. Panel (b): distributions of the deviation from $\mathrm{mean}[\mathcal{J}_\rho^{\hat{T}}]$ for $\hat{T}=500$. The grey parabola is a Gaussian fit to the distribution for $\hat{T}=5000\nu$.}
	\label{fig:pdfs-sens-ks}
\end{figure}

Probability distributions of the gradient $\mathcal{J}^{\hat{T}}_\nu$ for typical periodic orbits of the KS system are reported in panel (a) of figure \ref{fig:pdfs-sens-ks}. Near the peak, the distributions can be approximated reasonably well by a Gaussian law. However, much higher/lower sensitivities are observed for a few orbits, resulting in a significant departure from normality and heavy tails. To characterize these tails more precisely, we find fifteen thousand more periodic orbits for $\hat{T}=500$ and show in panel (b) the probability distribution of the deviation from the mean of the distribution in panel (a), for this reference period. The tails are well described by a power-law distribution of the form $p(x) = x^{-n}$ with exponent $n=3$. This structure is an inherent feature of the problem and not a numerical artefact depending, for instance, on the resolution.

\begin{figure}[htbp]
	\centering
	\includegraphics[width=0.49\textwidth]{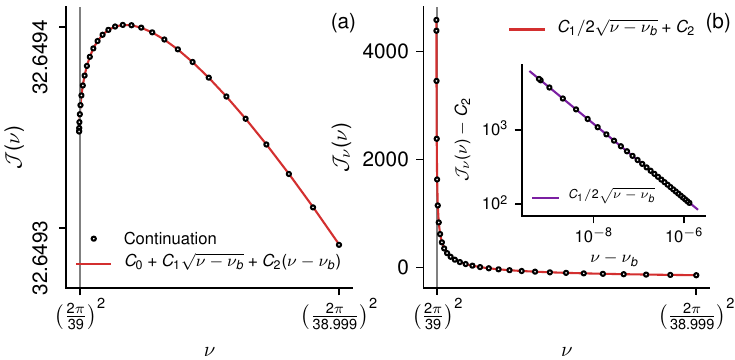}
	\caption{Panel (a): period average of an orbit with $\hat{T}=500$ with large positive gradient $\mathcal{J}_{\nu} \simeq 1572.27$, continued over the parameter $\nu$. Panel (b): sensitivity of the period average for the same orbit. Open circles denote data from the continuation. The red curve is the model (\ref{eq:J_alpha}) fitted to the data. The smaller inset shows the same quantity on shifted coordinates, in a bi-logarithmic plot.}
	\label{fig:continuation-ks}
\end{figure}

As illustrated in section \ref{sec:forward-sensitivity-floquet}, large sensitivities can be directly associated to bifurcations. This is illustrated in figure \ref{fig:continuation-ks}, where we report the continuation analysis of an orbit at $\hat{T} = 500$ with large gradient $\mathcal{J}_{\nu} \simeq 1572.27$. The average energy density $\mathcal{J}(\nu)$ and its gradient $\mathcal{J}_{\nu}(\nu)$ are reported as a function of the bifurcation parameter $\nu$ in panels (a) and (b), respectively. 
Near the bifurcation point, denoted by $\nu_b$, the period average is well described by the functional form
\begin{equation}\label{eq:J}
\mathcal{J}(\nu) \simeq C_0 + C_1\sqrt{\nu - \nu_b} + C_2(\nu - \nu_b),\quad \nu > \nu_b,
\end{equation}
where the square root term is typical in normal forms of bifurcations for periodic orbits \cite{Guckenheimer:up}. The gradient is then 
\begin{equation}\label{eq:J_alpha}
\mathcal{J}_\nu(\nu) \simeq \frac{C_1}{2\sqrt{\nu - \nu_b}} + C_2
\end{equation}
and approaches infinity at $\nu_b$ as $1/\sqrt{\nu}$. Fitting the data in panel $(b)$ to the model (\ref{eq:J_alpha}) shows that the constant $C_2$, the gradient measured sufficiently far away from the bifurcation point, is about $-246.65$, in line with the high probability region of the distributions in figure \ref{fig:pdfs-sens-ks}-(a). 

 

The functional form (\ref{eq:J}) is sufficient to explain the structure of the tails in figure \ref{fig:pdfs-sens-ks}. To this end, assume that periodic orbits appear in bifurcations at critical values $\nu_b$ as $\nu$ is increased, as expressed by (\ref{eq:J}).
Focusing at a given $\hat{T}$, assume also that the number of periodic orbits is large, so that $\mathcal{J}(\nu)$ can be thought of as a random variable, with the coefficients $C_0, C_1, C_2$ and the bifurcation point $\nu_b$ being random variables with values differing from orbit to orbit. The gradient $\mathcal{J}_\nu(\nu)$ is then also a random variable that can take arbitrarily large values if $\nu - \nu_b$ is small. Now, the probability that the gradient $\mathcal{J}_\nu(\nu)$ is less than some large positive constant $x$ can be expressed by introducing the cumulative distribution function $P_{\mathcal{J}_\nu}(x)$, defined as
\begin{align}\label{eq:J_alpha_cdf}
	P_{\mathcal{J}_\nu}(x) =\;& \mathrm{prob}\left[\mathcal{J}_\nu(\nu) < x\right] \nonumber \\ =\;& 1 - \mathrm{prob}\left[\mathcal{J}_\nu(\nu) > x\right] \nonumber \\ =\;& 1 - \mathrm{prob}\left[ \nu - \nu_b < (C_1/2x)^2\right],
\end{align}
where we have used the definition (\ref{eq:J_alpha}) and neglected $C_2$, since $x\gg 1$, to develop the algebra in the last step. The probability in the third line can be equivalently interpreted as the probability that bifurcation points are closer to the reference value than a distance $(C_1/2x)^2$. Assuming the points $\nu_b$ not to be preferentially distributed on the real line near $\nu$, this probability is then $cx^{-2}$ for some constant $c$. In other words, the larger $x$, the less likely is that a periodic orbit bifurcates near $\nu$. Hence, the cumulative distribution of $P_{\mathcal{J}_\nu}(x)$ must, asymptotically for large $x$, obey 
\begin{equation}\label{eq:J_alpha_cdf2}
	P_{\mathcal{J}_\nu}(x) = 1 - c x^{-2}.
\end{equation}
The probability distribution of the gradient $\mathcal{J}_\nu$ can then be obtained by differentiating the cumulative distribution with respect to its argument, leading to the power-law
\begin{equation}\label{eq:J_alpha_pdf}
	p(x) = c x^{-3},\quad x\gg 1,
\end{equation}
the behaviour observed in figure \ref{fig:pdfs-sens-lorenz}. More generally, sampling functions that have poles of the form $(\nu - \nu_b)^\gamma$, produces probability distributions with power-law tails of the form $p(x) \simeq x^{-n}$, $x\gg 1$, with exponent $n = (\gamma-1)/\gamma$, leading to $n = 3$ for the present case with $\gamma = -1/2$ of equation (\ref{eq:J_alpha}).
\begin{figure}[htbp]
	\centering
	\includegraphics[width=0.49\textwidth]{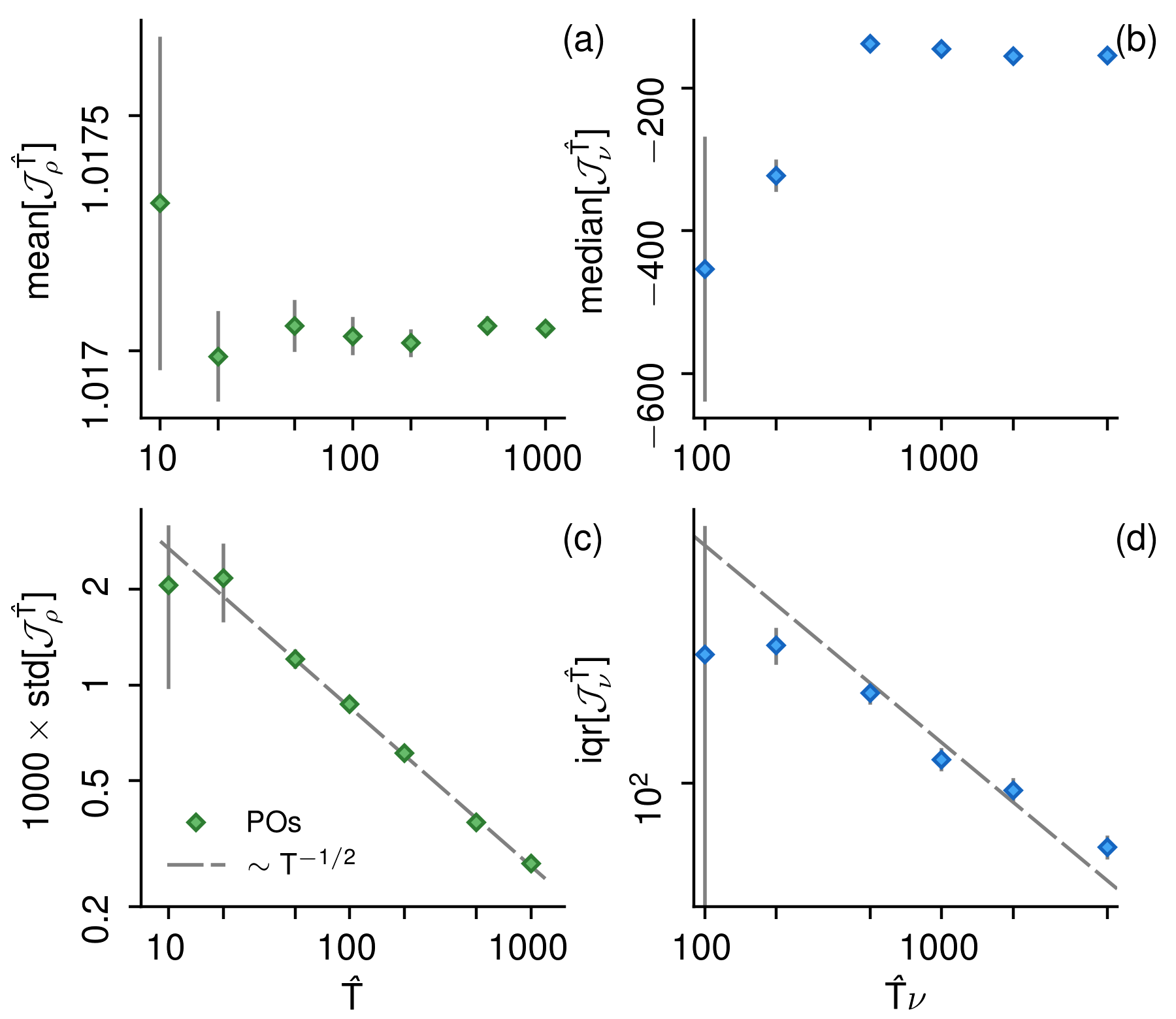}
	\caption{Dependence of the average, panels (a) and (b), and standard deviation, panels (c) and (d), of the gradient $\mathcal{J}^{\hat{T}}_\rho$, for the Lorenz equations (left panels), and $\mathcal{J}^{\hat{T}}_\nu$ for the KS system (right panel), on the reference period $\hat{T}$.  For the KS equations, the median and interquartile range are used, denoted as $\mathrm{median}[\cdot]$ and $\mathrm{iqr}[\cdot]$. Error bars for the KS system are estimated using a bootstrapping technique.}
	\label{fig:std-mean-sens} 
\end{figure}

We now come to the central result of this paper and examine the sensitivity of typical periodic orbits as a function of the reference period. For distributions with power-law tails of the form $p(x) \simeq x^{-n}$, central moments of order $m$ are undefined for $m\ge n-1$. For the KS system, with $n=3$, while the mean sensitivity across periodic orbits is defined (although convergence is weak), the standard deviation is not. Hence, for this system, we use the median and interquartile range, as measures for the localisation and variability of gradients, respectively Results are shown in figure \ref{fig:std-mean-sens}. Panels (a) and (b) show the mean and median gradient for the Lorenz and KS systems, respectively. Panels (c) and (d) show the standard deviation and interquartile range of the gradient. We observe that, as $\hat{T}\rightarrow\infty$ the mean/median converges to a value that is approximately $\mathcal{J}_\rho$ = 1.017 and $\mathcal{J}_\nu = -155$, for the Lorenz and KS systems, respectively. Given the bound (\ref{eq:upper-bound-on-solution}), the convergence of sensitivities is consistent with the convergence of the Floquet exponents of figure \ref{fig:std-mean-floquet}. However, for short cycles, the sensitivity of periodic orbits can be, in average terms over the inventory of available orbits, remarkably different to that of long cycles. This behaviour is more pronounced for the KS system. The standard deviation and interquartile range of the sensitivity  follow the same asymptotic behaviour of the period averages and decay asymptotically as ${\hat{T}}^{-1/2}$, indicating that the corresponding probability distributions localise around the averages of panels (a) and (b). An important remark is that sensitivity computations using periodic orbits do not suffer from shadowing errors displayed by shadowing methods applied to chaotic trajectories, e.g. the Least-Squares Shadowing \cite{Wang:2014bt} and Periodic Shadowing algorithms \cite{Lasagna:2018uo}. For such algorithms, convergence proofs have been offered that suggest that the standard deviation of sensitivity calculations on hyperbolic systems should first decay as ${\hat{T}}^{-1}$ as a result of the approximations of the exact shadowing direction involved in these algorithms.

Overall, these results show that the sensitivity computed from typical long periodic orbits found in computations converges to a well defined value as the period increases. This is not to say that all long orbits can be considered good proxies, as the tails of figure \ref{fig:pdfs-sens-ks} demonstrate, but rather that sensitivities computed from longer orbits found in computation are likely to be closer to the asymptotic value. This asymptotic value of the sensitivity is now compared with the response of long-time averages to finite-amplitude parameter perturbations using long chaotic simulations. Carefully conducted numerical approximations of the gradient using a finite-difference formula (see Ref.~[\onlinecite{Lasagna:2018uo}] for details) show that the response of the average of $u_3$ to perturbation of $\rho$ in the Lorenz equations is approximately $\mathcal{J}_\nu \simeq 1.002$, well below the asymptotic value from long periodic orbits. We remark that this difference is likely not a bias arising from \textcolor{black}{using} periodic orbits. In fact, the same difference has been previously observed using other shadowing algorithms applied to chaotic trajectories \cite{Lasagna:2018uo}. Numerical evidence has been provided \cite{Lucarini:2009jw} suggesting that the Lorenz equations have a linear response to perturbations of the parameter $\rho$, despite not being hyperbolic (it is a singularly hyperbolic system in the terminology of Ref.~[\onlinecite{Morales:1999iu}]). For this system, it has also been speculated \cite{Reick:2002co} that some observables might vary continuously with parameters and that the bifurcating orbits have very long period and their effect of the invariant measure is negligible. How to reconcile the existence of a linear response with the difference we observe between the prediction of shadowing methods and the actual response of the system is a question that deserves further analysis. 

\begin{figure}[htbp]
	\centering
	\includegraphics[width=0.49\textwidth]{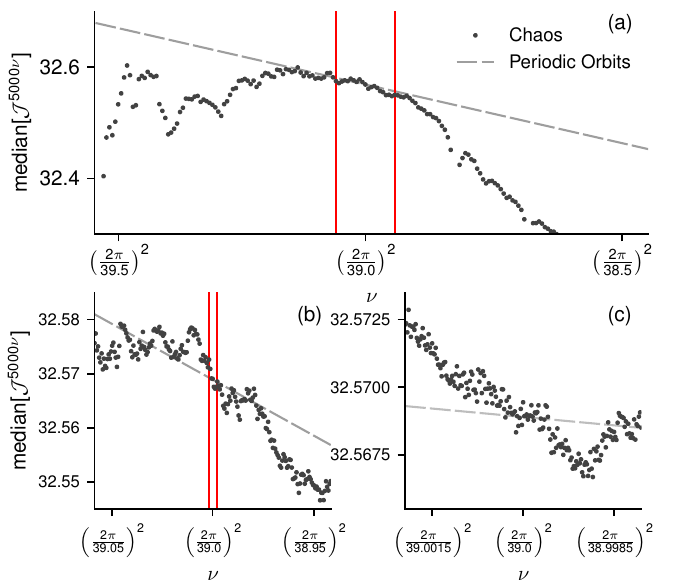}
	\caption{Long-time mean of the energy density as a function of $\nu$. Data points denote the median time average across thousand of simulations from different initial conditions, with averaging time $T=5000\nu$. The dashed line represents the slope predicted by periodic orbits with reference period $\hat{T} = 5000\nu$ \textcolor{black}{at the reference diffusivity $\nu = (2\pi/39)^2$}. Panels (b) and (c) focus on the area between the two red vertical lines in panels (a) and (b), respectively. }
	\label{fig:long-term-average} 
\end{figure}

For the KS system, we show in figure \ref{fig:long-term-average} how the long-time averaged energy density varies with the diffusivity. The data points are obtained by first computing averages over tens of thousands of long independent segments of length $T=5000\nu$, and then reporting the median value, which is more robust to outliers arising from initial conditions leading to a non-chaotic state. Using a bootstrapping technique we have also computed the standard error on the median, which is typically smaller than the symbol size in the figure, and it is thus not shown. We use this approach, instead of reporting the time average of one long chaotic computation, as it provides a measure of the accuracy of long-time average estimate. Panel (b) and (c) focus near the reference diffusivity $\nu = (2\pi/39)^2$ in the area spanned by the vertical lines in panels (a) and (b), respectively.
The dashed line represents the asymptotic gradient $\mathcal{J}_\nu$ from periodic orbits. The system clearly lacks a linear response, in the sense that the limit
\begin{equation}
	\lim_{\delta \nu \rightarrow 0} \frac{\mathcal{J}^\infty(\nu + \delta\nu) - \mathcal{J}^\infty(\nu) }{\delta \nu},
\end{equation}
is not defined, as the response of the system is not proportional to the perturbation in the parameter \cite{Reick:2002co,Farmer:1985iv,Ershov:1993el}, at any scale. As the distributions of sensitivities in figure \ref{fig:pdfs-sens-ks} suggest, some orbits are always infinitesimally close to bifurcation and small parameter perturbations might induce abrupt changes in the structure of the attractor, making chaotic averages non-differentiable. In such conditions, the meaning of gradients obtained from linear methods, either on periodic or chaotic trajectories, is unclear.

\section{Conclusions}
In this paper we set out to address the question whether typical long periodic orbits found numerically may be used as accurate proxies for the sensitivity of the chaotic state to parameter perturbations. Our motivation to address this question arises from well known challenges in locating periodic orbits in fluid systems governed by the Navier-Stokes equations, and the consequential possibility that predictions of cycle averaging formulae using an incomplete set might be inaccurate. If the answer to the above question is affirmative, an heuristic strategy would be to spend available computational resources to locate one or few orbits, of sufficiently long period. Accurate sensitivity information from these orbits may then facilitate control and optimisation tasks.

Here, we have considered long periodic orbits of two low-dimensional chaotic systems, the Lorenz equations at standard parameters and a minimal-domain Kuramoto-Sivashinky system with dynamics restricted to the anti-symmetric subspace. We have built an inventory of thousands of periodic orbits with period up to two orders of magnitude larger than the shortest admissible cycle. This approach was not guided by the idea of obtaining an exhaustive hierarchy of cycles and a complete understanding of their properties. Rather, we aimed to examine in statistical terms the properties of typical long orbits. This analysis is naturally biased by the search process, but the bias leans precisely towards the direction required to answer our original question, i.e.~it favours orbits that can be found in practical computations.

One conclusion from this study is that period averages of long orbits appear to converge to the long time average of chaotic trajectories. Floquet exponents, being the period averages of the local rate of growth of infinitesimal perturbations, also exhibit the same behaviour. Hence, Floquet exponents of long orbits converge to the Lyapunov exponents calculated using standard methods. These results are interesting, but perhaps not so important from an operational point of view, since these quantities can be calculated directly from chaotic trajectories at a lower cost. The important result is that the sensitivity of period averages of typical periodic orbits also converges to a defined value as the period increases. This result is consistent with the convergence of the Floquet exponents, based on the relation between stability and sensitivity of orbit, established in this paper. Interestingly, the probability distribution of sensitivities from typical orbits can display power-law tails of the form $p(x) \simeq x^{-3}$. This result can be explained by using a statistical argument and classical normal forms of bifurcations of periodic orbits. In practice, this suggests that sensitivity information from orbits found numerically may occasionally be quite inaccurate.

Some open challenges are now considered. First, as also observed elsewhere with other shadowing algorithms \cite{Blonigan:2018gd,Lasagna:2018uo} the sensitivity of most typical long periodic orbits is not necessarily consistent with the response of the system to finite-amplitude parameter perturbations. In absence of differentiability, the meaning and value of sensitivities from periodic orbits (or other shadowing algorithms) is unclear and deserves further analysis. The conjecture is that for high-dimensional systems, where statistics behave as if a linear response existed \cite{Albers:2006ei,Ruelle:1999bm}, the ``thermodynamic limit'' of Ruelle \cite{Ruelle:1999bm}, may be invoked and a better consistency between sensitivities from shadowing methods and the response of the system might be observed \cite{Blonigan:2018gd, Lasagna:2018uo}. However, evidence in support of this conjecture is currently lacking. 

A second challenge is that the applicability of the ideas discussed in this paper to fluid systems governed by the Navier-Stokes remains unclear (see Ref.~\cite{Chandler:2013fi}). One major issue is that the increase of system dimension inevitably implies a decrease of good near recurrence events. As advocated in Ref.~\cite{vanVeen:2019bi} more robust search methods are required. A third potential issue is that the size of the linear systems arising in the Newton-Raphson search iterations grows linearly with the period $T$, regardless of the numerical method utilized, i.e.~either for global search methods \cite{Lan:2004ch,Lasagna:2017tz} or with multiple-shooting techniques \cite{Ascher:1994ty,Cvitanovic:2010gm}. The condition number of these problems grows with $T$, introducing errors in the Newton-Raphson corrections that might eventually prevent convergence. Hence, finding long periodic orbits might, eventually, prove too challenging. 

\appendix 




\section{Floquet exponents of long periodic orbits}\label{app:numerical-method-long-periodic orbits-floquet}
The algorithm used to compute Floquet exponents of long periodic orbits follows closely the approach originally introduced in Ref.~[\onlinecite{Ding:2014vl}]. For completeness, we describe in this appendix this algorithm and outline our novel contribution.

The algorithm exploits two fundamental facts. The first is that the Jacobian matrices (\ref{eq:jacobian-matrices}) obey the multiplicative property
\begin{equation}
	\mathbf{M}(t_2, t_0) = \mathbf{M}(t_2, t_1)\mathbf{M}(t_1, t_0),
\end{equation}
for any times $t_2 \ge t_1 \ge t_0$. Hence, the monodromy matrix $\mathbf{M}(T, 0)$ can be equivalently expressed as the product of $M$ short-time Jacobian matrices $\mathbf{M}_i = \mathbf{M}(t_i, t_{i-1})$, $i = 1, \ldots, M$, as
\begin{equation}\label{eq:product-of-jacobians}
	\displaystyle \mathbf{M}(T, 0) = \mathbf{M}_{M}\mathbf{M}_{M-1}\ldots\mathbf{M}_{1},
\end{equation}
for a partitioning of the interval $[0, T]$ into $M$ sub-intervals specified by times $0 \equiv t_0 > t_1 > \ldots > t_{M-1} > t_M \equiv T$. Note that the Jacobian matrices $\mathbf{M}_i$ obey the cyclic property $\mathbf{M}_{M+1} = \mathbf{M}_1$.

The second fact is that a well-conditioned eigenvalue revealing decomposition exists for products of matrices such as (\ref{eq:product-of-jacobians}). This is the \textit{periodic real Schur decomposition} \cite{BojanczykGV:ai4u6Z78,Hench:jz}, initially introduced in the context of Floquet analysis in Ref.~[\onlinecite{Lust:2001iy}] for the computation of the multipliers and more recently extended \cite{Ding:2014vl,Ding:2016iw} to compute the eigenfunctions. This decomposition consists in factorizing the short-time Jacobian matrices using a set of orthogonal matrices $\mathbf{Q}_i$, $i=1, \ldots, M$, satisfying the cyclic property $\mathbf{Q}_0 = \mathbf{Q}_M$, as  
\begin{equation}\label{eq:prsf}
	\mathbf{M}_i = \mathbf{Q}_i\mathbf{R}_i\mathbf{Q}_{i-1}^\top
\end{equation}
where the factors $\mathbf{R}_i, i=1, \ldots, M-1$ are upper triangular matrices and $\mathbf{R}_M$ is in {real Schur form}, a block upper-triangular matrix with either $1\times 1$ and $2\times 2$ blocks on the diagonal, in case the monodromy matrix possesses pairs of complex conjugate multipliers. 

Using these two facts, the monodromy matrix can be expressed in \textit{real Schur form} as
\begin{equation}\label{eq:rsf}
	\mathbf{M}(T, 0) = \mathbf{Q}_0 \mathbf{R}_M...\mathbf{R}_2\mathbf{R}_1 \mathbf{Q}_{0}^\top.
\end{equation}
The product $\mathbf{R}_M...\mathbf{R}_2\mathbf{R}_1$ and the monodromy matrix are unitarily similar and thus share the same spectrum of eigenvalues. However, because of the structure of the factors $\mathbf{R}_i$, obtaining the spectrum is a straightforward computation, since the spectrum of a block triangular matrix is the union of the spectra of the blocks. The structure of the block upper triangular factor $\mathbf{R}_M$ determines whether exponents are real or form complex conjugate pairs (see Ref.~[\onlinecite{Golub:2012wp}], Th. 7.4.1). For a $1\times 1$ block at location $(i, i)$, a real Floquet exponent can be obtained as
\begin{equation}\label{eq:exponents-from-multipliers-real}
	\lambda_i = \log(\mu_i)/T = \frac{1}{T}\log \prod_{j=1}^M [\mathbf{R}_j]_{ii} = \frac{1}{T}\sum_{j=1}^M \log [\mathbf{R}_j]_{ii}
\end{equation}
Computing the sum of the logarithms is recommended, as multiplication can quickly over/underflow before the logarithm is taken. For a $2\times 2$ block, a pair of complex conjugate exponents can be obtained with a bit more work by recursively multiplying all $2\times 2$ blocks of the factors $\mathbf{R}_j$ at location $(i, i+1)$, and accumulating the sum of the logarithms of scaling factors required to set the largest element in the partial products to have unitary magnitude. Overall, this algorithm only operates on well-conditioned short-time Jacobian matrices, instead of forming the monodromy matrix, and it is numerically robust.

The numerical algorithm required to obtain the factors $\mathbf{R}_i$ and $\mathbf{Q}_i$ in equation (\ref{eq:prsf}) from the short-time jacobian matrices is based on classical QR-based eigenvalue algorithms \cite{Kressner:2006dd,Golub:2012wp}. Developing a robust implementation is a lengthy and delicate task. In this paper, we have adopted a matrix-free algorithm introduced in Ref.~[\onlinecite{Ding:2014vl}], which is a specialization to periodic orbits of classical methods to compute Lyapunov exponents \cite{Benettin:1980ek}.  The approach only requires the action of these matrices on a set of tangent vectors and is thus computationally more efficient when only a handful of the leading Floquet exponents is required. The algorithm is simpler to implement, does not require advanced linear algebra technique and only requires minor modifications to an existing time-stepper code for the linearised equations. 



The algorithm is iterative and we denote quantities at iteration $k$, with a  superscript $(k)$. First, a set of $m$ linearly independent tangent vectors is defined, where $m$ is the number of desired exponents. For notational convenience, we arrange them as the columns of matrix $\hat{\mathbf{Q}}_0^{k}\in\mathbb{R}^{n, m}$, where we use a hat to denote a matrix with reduced dimensions. The period is divided into $M$ sub-intervals, which need not have the same size. In each sub-interval, the columns of $\hat{\mathbf{Q}}_0^{k}$ are: 1) propagated forward in time using a linearised time-stepping solver and 2) re-orthogonalized in place using a Gram-Schmidt procedure. These two steps are formally equivalent to computing
\begin{equation}
	\mathbf{M}_i \hat{\mathbf{Q}}_{i-1}^{k} = \hat{\mathbf{Q}}_{i}^{k}\hat{\mathbf{R}}_{i}^{k}\quad i = 1, \ldots, M,
\end{equation}
which is akin to (\ref{eq:prsf}). Note that forming $\mathbf{M}_i$ is not necessary, only its action on the columns of $\hat{\mathbf{Q}}_{i-1}^{k}$ is required, making the approach suitable for PDE problems. The triangular factors $\hat{\mathbf{R}}_{i}^{k} \in \mathbb{R}^{m\times m}$ from the orthogonalization are stored for later processing. The time between subsequent re-orthogonalizations depends on the expanding/contracting characteristics of the tangent space and should be chosen such that the columns of $\hat{\mathbf{Q}}_i^{k}$ remain numerically linearly independent. 

After the last sub-interval, the iterations are restarted by setting $\hat{\mathbf{Q}}_{0}^{k+1} = \hat{\mathbf{Q}}_{M}^{k}$ and after $k$ iterations the monodromy matrix is formally equivalent to
\begin{equation}\label{eq:schur-after}
	\mathbf{M}(T, 0) = \hat{\mathbf{Q}}_{0}^{k+1} \hat{\mathbf{R}}_{M}^{k}\ldots\hat{\mathbf{R}}_{2}^{k}\hat{\mathbf{R}}_{1}^{k} \hat{\mathbf{Q}}_{0}^{k\top}.
\end{equation}
If the first $m$ Floquet multipliers are all real and distinct, the columns of $\hat{\mathbf{Q}}_{0}^{k+1}$ converge geometrically to a basis for the subspace spanned by the leading $m$ Floquet eigenvectors. Hence, the difference $\|\hat{\mathbf{Q}}_{0}^{k+1} - \hat{\mathbf{Q}}_{0}^{k}\|$ converges to zero in any norm and (\ref{eq:schur-after}) is the real Schur form of the monodromy matrix, as in equation (\ref{eq:rsf}). In fact, this iteration procedure is a form of subspace iteration \cite{Saad:2011tu} (also known as orthogonal iteration \cite{Golub:2012wp}, and referred to as simultaneous iteration in Ref.~[\onlinecite{Ding:2014vl}]). The leading $m$ Floquet exponents can then be obtained from the diagonals of the factors $\hat{\mathbf{R}}_i$, as discussed.

\begin{figure*}[htbp]
	\centering
	\includegraphics[width=0.99\textwidth]{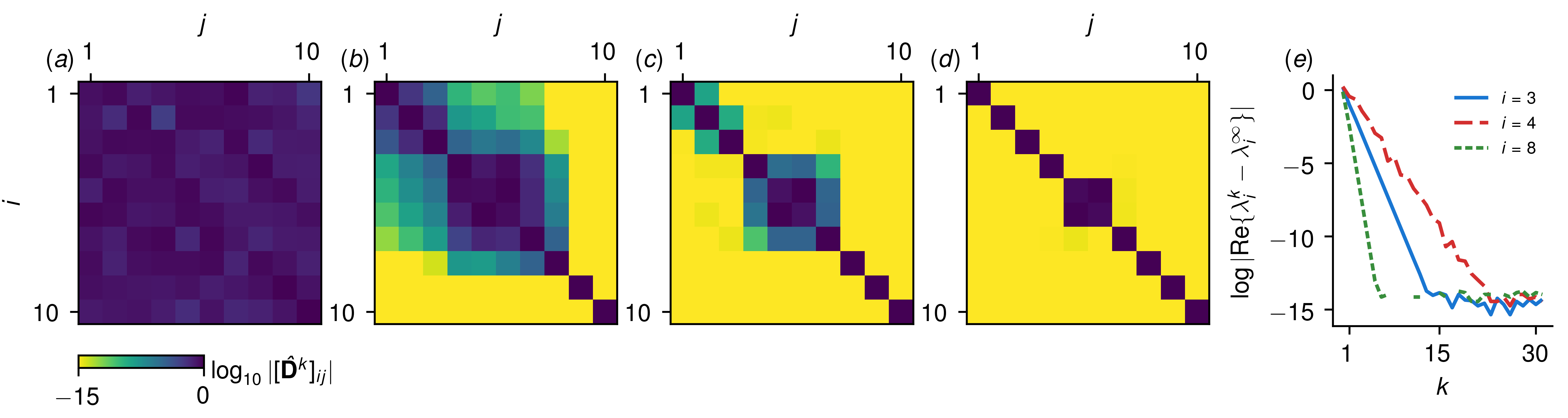}
	\caption{The base ten logarithm of the entries of the rotation matrix $\hat{\mathbf{D}}_k$ at iterations $k=1, 3, 10$ and 33, panels (a-d). Convergence of error on the estimate of a few selected Floquet exponents, panel (e).}
	\label{fig:floquet-algorithm} 
\end{figure*}

A simple adjustment of this approach can be introduced when some of the multipliers form complex conjugate pairs. The iterations still converge, in the sense that the subspace spanned by $\hat{\mathbf{Q}}_{0}^{k}$ converges, but only the columns associated to real exponents converge individually \cite{Ding:2014vl}. The subspace spanned by a pair of columns of $\hat{\mathbf{Q}}_{0}^{k+1}$ corresponding by the space spanned by the Floquet eigenvector associated to complex conjugate multipliers also converges, but at every iteration the two columns are rotated by an angle in the subspace they span. Hence, we introduce a rotation matrix $\hat{\mathbf{D}}^{k}$ such that
\begin{equation}\label{eq:D-rotation} 
\hat{\mathbf{Q}}_{0}^{k+1} = \hat{\mathbf{Q}}_{0}^{k} \hat{\mathbf{D}}^{k}.
\end{equation}
For large $k$, this rotation converges to a matrix that has the structure of the product of Givens rotation matrices, each rotating one pair of columns of $\hat{\mathbf{Q}}_{0}^{k}$ to the corresponding pair of $\hat{\mathbf{Q}}_{0}^{k+1}$. With this modification, the product $\hat{\mathbf{D}}^k \hat{\mathbf{R}}^k_M$ converges to a block upper triangular matrix and (\ref{eq:schur-after}) with (\ref{eq:D-rotation}) is formally equivalent to (\ref{eq:rsf}).

To the best of the author's understanding, a procedure to compute this rotation matrix was not proposed Ref.~[\onlinecite{Ding:2014vl}], which focused instead on using iterative QR-based algorithms. The new contribution of this appendix is a simple strategy to obtain it. The rotation $\hat{\mathbf{D}}^{k}$ can be found as the solution of the \textit{orthogonal Procrustes problem} \cite{Golub:2012wp} 
\begin{equation}
	\underset{\mathbf{D}^k}{\mathrm{argmin}}\;\; \|\hat{\mathbf{Q}}_{0}^{k+1} - \hat{\mathbf{Q}}_{0}^{k} \mathbf{D}^{k} \|_{\mathrm{F}} = \mathbf{U}^{k} \mathbf{V}^{k\top},
\end{equation}
where $\|\cdot\|_\mathrm{F}$ is the Frobenius norm and where the two matrices at the right hand side are obtained from the Singular Value Decomposition 
\begin{equation}
\hat{\mathbf{Q}}_{0}^{k\top}\hat{\mathbf{Q}}_{0}^{k+1} = \mathbf{U}^{k} \mathbf{\Sigma}^{k} \mathbf{V}^{k\top}.
\end{equation}

In our implementation, we compute the rotation $\hat{\mathbf{D}}^{k}$ along the iterations and use a simple heuristic to detect pairs of complex conjugate eigenvectors, or inverse hyperbolic directions, when a diagonal entry is close to $-1$. We then monitor the maximum absolute difference between estimates of the Floquet exponents and stop the iterations when such difference is lower than a user-defined tolerance. Panels (a-d) of figure \ref{fig:floquet-algorithm} shows the progressive convergence of the rotation matrix $\hat{\mathbf{D}}_k$ for a calculation on the shortest periodic orbit of the KS system reported in figure \ref{fig:short-long}. Iteration $k=1, 3, 10$ and 33 are shown. Except for the fifth and sixth column, the columns of $\hat{\mathbf{Q}}_{0}^{k+1}$ converge to the columns of $\hat{\mathbf{Q}}_{0}^{k}$ and all Floquet multipliers are real and positive. Panel (e) shows the convergence of the error on the estimate of a few Floquet exponents.

In practice, we have found this method to be quite robust for the systems used in this paper, where Floquet exponents are typically well separated. We have observed that the number of iterations required for convergence decreases with the period, with exponents of the longest periodic orbits of the KS system requiring only three/four iterations to converge to machine accuracy. This can be attributed to the faster convergence of the columns of $\hat{\mathbf{Q}}^k_i$ to the subspace spanned by the leading Floquet eigenmodes, as the integration time is proportionally longer, following the same pattern of convergence of algorithms to compute Lyapunov exponents from chaotic trajectories \cite{Goldhirsch:1987hm}. We have therefore made no attempt at improving the convergence rate and computational cost by using shift and deflation techniques that are customarily used in state-of-the-art implementations of eigenvalue algorithms \cite{Golub:2012wp,Kressner:2006dd}.

\bibliography{library}
 
\end{document}